\documentclass[lettersize,journal]{IEEEtran}
\usepackage{graphicx,comment}
\usepackage[colorlinks=true, allcolors=blue]{hyperref}
\usepackage{soul}
\usepackage{amsmath}
\usepackage{amssymb}
\usepackage{algpseudocode}
\usepackage{enumitem}
\usepackage{mathrsfs}
\usepackage[table]{xcolor}
\usepackage{caption}
\usepackage{subcaption}
\newcommand{\red}[1]{\textcolor{red}{#1}}
\newcommand{\blue}[1]{\textcolor{blue}{#1}}

\usepackage{bbm}
\usepackage{algorithm}
\usepackage{tabularx}

\def\BibTeX{{\rm B\kern-.05em{\sc i\kern-.025em b}\kern-.08em
    T\kern-.1667em\lower.7ex\hbox{E}\kern-.125emX}}
\usepackage{fancyhdr}

\makeatletter
\let\ps@IEEEtitlepagestyle
\makeatother

\ifCLASSOPTIONcompsoc
  \usepackage[nocompress]{cite}
\else
  \usepackage{cite}
\fi

\ifCLASSINFOpdf
 
\else
  
\fi

\begin{document}

\title{Radio Resource Management for the Uplink of Hybrid Beamforming  Systems}

\author{Yuan~Quan, Haseen~Rahman, and~Catherine~Rosenberg,~\IEEEmembership{Fellow,~IEEE}
        
\thanks{Part of this paper was presented in {\it IEEE ICC}, Montreal, Canada, June 2025 \cite{ULConf}.

Y. Quan, H. Rahman and C. Rosenberg are with the Department of Electrical and Computer
Engineering, University of Waterloo, Waterloo, ON N2L 3G1, Canada (e-mail:
yuan.quan@uwaterloo.ca; hrcheriyandilakath@uwaterloo.ca; cath@uwaterloo.ca).}}


\maketitle

\begin{abstract}

This paper studies radio resource management (RRM) for the uplink of a multi-channel cellular system with hybrid beamforming based on analog beamforming using predefined codebooks and zero-forcing digital beamforming. We first formulate a per-time slot joint RRM optimization problem, which includes beam selection, user selection, power allocation, modulation and coding scheme selection, and digital beamforming. A per-time slot formulation of the RRM problem is necessary because the power budget of a user equipment (UE) needs to be allocated per time slot to its assigned channels which are not known a priori, and because we consider the case where the number of radio frequency chains is not large enough to select all possible analog beams, thereby requiring per-slot beam selection. This problem can be solved for at most a few UEs because the number of variables grows exponentially with the number of UEs. 
In order to obtain results with more UEs, we propose an offline heuristic that reduces the runtime to obtain results by two orders of magnitude, while achieving performance close to the joint optimization. This offline heuristic allows us to obtain engineering insights on the impact of different system parameters as well as a target performance that we use to validate the low-complexity online heuristic that we propose.



\end{abstract}
\begin{IEEEkeywords}
 Radio resource management, uplink, hybrid beamforming, power allocation, massive MIMO, 5G and beyond.
\end{IEEEkeywords}

\IEEEdisplaynontitleabstractindextext

\IEEEpeerreviewmaketitle

\section{Introduction}\label{sec:introduction}

Beamforming (BF) and massive multi-input multi-output (MIMO) are two of the key technologies that enable high capacity and spectrum-efficient communications in 5G and beyond systems. Fully digital BF (FDBF) provides the maximum flexibility and performance, but its deployment is costly since it requires one radio frequency (RF) chain per antenna \cite{MIMO1}. Instead, hybrid beamforming (HBF) combines digital BF (DBF) with analog BF (ABF) implemented through phase shifters. HBF can be implemented with a number of RF chains much smaller than the number of antennas while enabling multi-stream data transmission within the same channel \cite{hybrid-survey}. However, the ABF component imposes practical constraints that are not present in FDBF, making radio resource management (RRM) in HBF systems challenging.

Most of the RRM research on HBF systems has focused on the downlink (DL), while the uplink (UL) RRM has received less attention despite the growing recognition that the UL can be a bottleneck affecting the perceived quality of experience \cite{strinati}.
In this paper, we consider the UL of a time division duplexing (TDD) and orthogonal frequency division multiple access (OFDMA) based single cell using HBF. The bandwidth and time are organized into physical resource blocks (PRBs), each corresponding to a channel in a time slot. Some time slots are allocated to the DL and others to the UL. RRM on the UL inherently differs from RRM on the DL because, on the UL,  the power budget, that belongs to a user equipment (UE) cannot be allocated to PRBs within a time slot before knowing on which PRBs the UE will be transmitting, which means that power allocation and user selection are inter-dependent \cite{DL,ECSIXiaomeng}. Consequently, designing RRM solutions tailored to the unique constraints of the UL are essential.

\begin{figure}
    \centering
    \includegraphics[width=80mm]{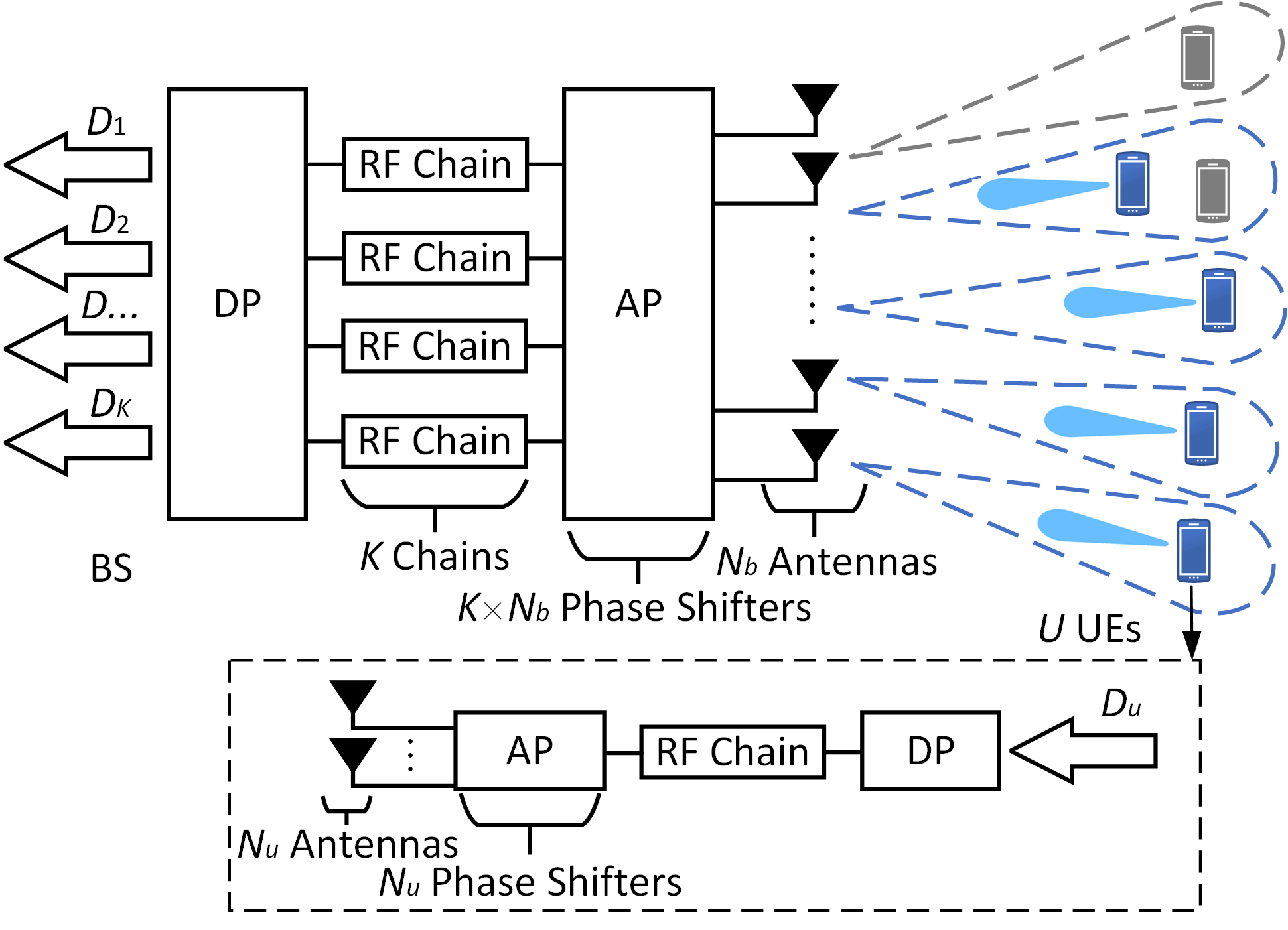}
    \caption{HBF architecture at the BS and UE and the transmission in one PRB (for illustration purposes $K=4$ and $U=6$).
    }
    \label{fHBF}
\end{figure}

We consider an HBF system comprising one base station (BS) and $U$ UEs. The BS employs a fully connected architecture with $K$ RF chains, where each RF chain is connected to every antenna through an independent phase shifter. We do not make any assumption on the relative value of $K$ with respect to $U$. Each UE is equipped with a single RF chain connected with all its antennas via independent phase shifters, as shown in Fig.~\ref{fHBF} \cite{DL}. Contrary to the scarce uplink literature that focuses almost exclusively on a single-channel case, we consider the multi-channel case, which is a challenging but more practical scenario, as will be discussed below.

The system selects analog beams from predefined BS and UE codebooks of large size (e.g., generally larger than the number of RF chains at the BS and the UE, respectively) \cite{CE1}.
During beam alignment (BA) \cite{3GPPBM},  each UE is matched with one or multiple BS beams and a single UE beam.  In one time slot, the BS can select, for data transmission, at most as many analog beams (among the matched ones) as the number of RF chains at the BS. The set of selected analog beams remains fixed within a time slot, since the phase shifters are shared across all channels. Hence, there is a need for beam selection if the number of distinct matched beams is larger than the number of RF chains. If a selected beam is matched to multiple UEs, more than one UE can transmit using that beam in the time slot (on different subchannels). ABF is thus achieved through beam selection, followed by per-channel user selection and DBF based on the selected analog beams to mitigate inter-beam interference.

Although beam selection is fixed across channels within a time slot, different UEs can be assigned to a BS beam in different channels as long as they have been matched to that beam. Hence, user selection is performed on a channel basis during RRM, given the selected beams (i.e., beam selection and user selection are coupled).

Another key objective is to evaluate the impact of the number of RF chains on performance. In practice, the number of RF chains is often much smaller than the number of UEs due to hardware limitations. However, this practical case has not been explored in existing UL studies (e.g., \cite{UL2_LCHB,UL3_RLCH,UL5_HPCD,UL4_HBDU,UL1_LCMR,UL7_JUAH}), which have focused on the case where the number of RF chains is no smaller than the number of UEs. We will see that the case where the number of RF chains is smaller than the number of UEs is more challenging to study. Our formulations and solutions apply to both cases. The detailed comparison and analysis of existing research are presented in Section~\ref{sec:literature}.



Our research focuses on the RRM for the uplink of the system described above, given that beam alignment and channel estimation have been performed. The RRM is time slot-based and comprises the following interdependent processes:
\begin{enumerate}[leftmargin=*]
\item \textit{Per time slot beam selection}: 
When the number of RF chains is no smaller than the number of UEs,  each UE can have its preferred matched BS beam  (the one offering the highest average channel gain over PRBs in the time slot) selected all the time and hence there is no need for beam selection. When the number of RF chains is smaller than the number of UEs, \emph{per time slot beam selection}  is necessary whenever the cardinality of the set containing the preferred matched BS beams for each UE is larger than the number of RF chains (which will happen often if the users are not clumped together). In this case, a subset of beams, with cardinality equal to the number of RF chains, must be selected from the set of all matched BS beams and must remain fixed for all the PRBs of the time slot. 
\item \textit{Per PRB user selection}: Given a selected beam set, per-PRB user selection has to be performed, for all selected BS beams that have been matched with multiple UEs. At most one UE will be selected from the UEs matched with this beam in each PRB. This is illustrated in Fig.~\ref{fHBF} where the blue beams and UEs are active and the gray beam and UEs are inactive. This creates an intrinsic coupling between beam selection and user selection \cite{DL}. 
\item \textit{Per user power allocation (PA)}: RRM on the UL has to be performed on a time slot basis, not only because of beam selection but also for \emph{power allocation} where each UE allocates its power budget to its assigned PRBs within the time slot.
PA is inherently coupled with user selection, as a UE needs to know the assigned channels\footnote{Within a time slot $t$, assigning a channel $c$ is equivalent to assigning the PRB $(c,t)$ and we will use channel and PRB interchangeably.} before allocating power to them. The interdependence of beam selection, user selection, and PA presents a significant challenge when studying the RRM for the uplink of the system described above.
\item \textit{Per PRB digital BF}: To mitigate inter-beam interference, we apply zero-forcing (ZF) combining at the BS to the data streams of the selected UEs. 
\item \textit{Modulation and coding scheme (MCS) selection}: Given beam selection, user selection, PA and DBF, the signal-to-noise ratio (SNR) (there is no interference because of ZF DBF) can be estimated for each selected stream in each PRB and the appropriate MCS can finally be selected for data transmission \cite{DL}.
\end{enumerate}


Our aims are twofold: 1)  to analyze the impact  of different system parameters such as the number of RF chains and the number of BS beams matched per UE on system performance as well as of different PA strategies (e.g., optimized PA versus water filling); 2)  to develop a low-complexity online RRM solution that performs well. For this second goal, we need a target performance to validate our online solution.

Specifically, the main contributions of the paper are:
\begin{itemize}[leftmargin=*]
  \item We assess the performance of the uplink of the HBF system described above that operates on multiple channels and aims at providing proportional fairness (PF) to its UEs. The number of RF chains can take any value larger than one,  and each UE matches with more than one BS analog beams during beam alignment. To do so, we formulate a joint RRM optimization problem, over a time slot, that includes beam selection, user selection, PA, DBF, and MCS selection. 
  \item This joint optimization problem is a large mixed-integer nonlinear programming (MINLP) problem. Although learning-based approaches have been employed to address MINLP problems \cite{ML1,ML2}, the optimality of such solutions remains unclear. Instead, We apply relaxation and approximation techniques to transform it into an upper-bounding linear programming (LP) problem. We then show how to derive a quasi-optimal feasible solution from that LP problem.
  \item Obtaining this quasi-optimal feasible solution is computationally intensive due to the large size of the problem and hence it can only be done for small systems (i.e., no more than a few UEs and RF chains). To enable system planning with a larger number of UEs and RF chains, we propose an offline  heuristic based on four steps, the two main steps being Step~1 and Step~3.  Step~1  optimizes user selection/PA performed per beam \emph{in isolation} (while ignoring the other beams), considering all users that have matched with this beam. Step~3 is a ``load-aware'' beam selection strategy where the $K$ beams with the largest loads are selected. We define the load of a beam as the weighted sum rate (WSR) obtained by performing user selection/PA  in isolation. The weights in the WSR are there to provide proportional fairness.  There are two other steps for that heuristic. Step~2 is necessary if the number of matched BS beams per UE is larger than one, since a user might be scheduled in the same PRB on multiple beams, which is not possible. We propose a simple way to remove these conflicts and modify results from Step~1 correspondingly. Given beam selection and user selection, Step~4 performs DBF per PRB over all selected beams for the UEs selected in that PRB and redo an optimized PA. This heuristic can achieve at least 93\% of the quasi-optimal performance while reducing the computation time to obtain meaningful results (for different numbers of RF chains and UEs)  from months to hours. Our formulations and solutions are independent of the frequency band on which the system operates.
  \item We obtain insights into the impact of system parameters from the offline optimization and the heuristic results in a mmWave setting. We analyze the impact of the number of BS beams matched per UE during BA, and of the number of RF chains at the BS. Results indicate that, using one matched BS beam per UE and a number of RF chains at the BS much smaller than the number of antennas, strikes a balance between performance, complexity and cost, serving as a guideline for deployment strategies and parameters.
  \item Finally, our online heuristic is inspired by the offline one where we simplify Steps~1 and 4. We propose a per-beam scheduler (PBS) to replace Step~1, i.e., the per-beam user selection and PA optimization used in the offline heuristic. The proposed PBS improves upon the state-of-the-art solution for a single cell scheduler proposed in \cite{PerBeam} by up to 22\%. Finally, in Step~4, water filling is used for PA. This online heuristic achieves at least 92\% of the performance of the offline heuristic in a mmWave setting.
\end{itemize}

The rest of the paper is organized as follows. Section~\ref{sec:literature} reviews the related work. Section~\ref{sec:sysmodel} describes the system model.  In Section~\ref{sec:optimization}, we formulate and solve the joint RRM optimization problem. Section~\ref{sec:heu} provides the offline and online heuristic RRM schemes. Section~\ref{sec:results} presents the numerical evaluation of our proposed optimization and heuristic solutions in a mmWave setting.  Section~\ref{sec:conclusion} concludes the paper. 

\section{Related Work} \label{sec:literature}














\begin{table}
\centering

\caption{Main system characteristics in the HBF UL references (Eff.: effective, CH: channel, OPT/O: optimized, WF: water filling, Pre-CBs: predefined codebooks, FN: fairness, NI: non-iterative solution).}
\footnotesize
\setlength{\tabcolsep}{1.6mm}{
\begin{tabular}{|c|c|c|c|c|c|c|c|c|c|} 

\hline

\rowcolor[HTML]{D3D3D3} & CSI & CH & $K < U$ & PA & ABF & DBF & FN & NI\\
\hline

\cellcolor[HTML]{D3D3D3} \cite{UL2_LCHB} & Full & $\geq 1$ & $\times$ & Equal & Online & MMSE & $\times$ & $\times$ \\
\hline

\cellcolor[HTML]{D3D3D3} \cite{UL3_RLCH} & Full & 1 & $\times$ & $\times$ & Online & MMSE & $\times$ & \checkmark \\
\hline

\cellcolor[HTML]{D3D3D3} \cite{UL5_HPCD} & Full & 1 & $\times$ & $\times$ & Online & MMSE & $\times$ & $\times$ \\
\hline

\cellcolor[HTML]{D3D3D3} \cite{UL4_HBDU} & Full & 1 & $\times$ & $\times$ & Pre-CBs & MMSE & $\times$ & \checkmark \\
\hline

\cellcolor[HTML]{D3D3D3} \cite{UL1_LCMR} & Eff. & 1 & $\times$ & $\times$ & Pre-CBs & ZF & $\times$ & $\times$ \\
\hline

\cellcolor[HTML]{D3D3D3} \cite{UL7_JUAH} & Eff. & 1 & $\times$ & $\times$ & Pre-CBs & OPT & PF & $\times$ \\
\hline

\cellcolor[HTML]{D3D3D3} Ours & Eff. & $\geq 1$ & \checkmark & O/WF & Pre-CBs & ZF & PF & \checkmark \\
\hline

\end{tabular}}

\label{ULRefTab}
\end{table}

In the introduction, we briefly noted that existing UL studies have not addressed the case $K<U$ in a multi-channel setting. In this section, we provide a detailed analysis and comparison of prior work related to UL RRM for HBF systems. Few studies have examined RRM for the uplink of HBF systems. Table~\ref{ULRefTab} summarizes related studies along with their system characteristics. The first thing to note is that, except for \cite{UL2_LCHB}, none of the papers work on multiple channels and hence they cannot address the PA and the beam selection correctly. \cite{UL2_LCHB}  assumes $K > U$, thus there is no need for beam or user selections.  Secondly, \cite{UL2_LCHB}, \cite{UL3_RLCH} and \cite{UL5_HPCD} are not based on predefined codebooks. Instead, they generate analog beams during RRM, which adds complexity and requires full channel state information (CSI). While \cite{UL4_HBDU} selects beams from predefined codebooks, it uses full CSI  for user selection. In contrast, our study uses  HBF based on predefined codebooks where all RRM processes use effective channels instead of full CSI, reducing both complexity and overhead as in \cite{UL1_LCMR} and \cite{UL7_JUAH}. Specifically, using effective CSI instead of full CSI can reduce the CSI dimension for each UE from an $N_u \times N_b$ matrix to a $U \times 1$ scalar. However, these two papers assume a single-channel scenario with enough RF chains to select all matched BS beams so that beam selection is not necessary, thereby bypassing one of the couplings that our work addresses. All the papers assume a rate function (i.e., one that maps an SINR into a rate) based on Shannon while we use a more practical (but also a more challenging) one based on a finite set of MCSs \cite{DL,MCS,MCS3}. Finally, only \cite{UL7_JUAH} from all the cited papers take fairness into account. We integrate fairness into our objective function, following the approach in \cite{UL7_JUAH,PF1,PF2}. 

One of our goals is to find an effective PA online solution. Per-time slot PA is not considered in \cite{UL3_RLCH,UL4_HBDU,UL5_HPCD,UL7_JUAH,UL1_LCMR} as they focus on single-channel scenarios. Meanwhile, \cite{UL2_LCHB} assumes all UEs are selected in every channel and equal PA across all channels within a time slot, which can be inefficient in some cases. In this paper, we start with a PA that optimizes PA based on practical MCS levels in our joint optimization and offline heuristic. We call it OPA. For our online heuristic, we adopt a low-complexity PA, which allocates each UE’s power across its assigned channels within a time slot after user selection using water filling (WF). We call it WFPA. We will evaluate the efficiency of WFPA by comparing the online heuristic to the offline heuristic.

Minimum mean square error (MMSE) or optimization-based DBF are employed in \cite{UL2_LCHB,UL3_RLCH,UL4_HBDU,UL5_HPCD,UL7_JUAH}. However, all these studies focus on scenarios where beam selection, user selection, and PA are not required because none of them considers $K \geq U$ with multiple channels, as summarized in Table~\ref{ULRefTab}.
From the best of our knowledge, there is no paper that has developed an optimal solution for a multi-channel system. Indeed, for a multi-channel system, MMSE DBF makes the joint optimization highly nonconvex because of interference between data streams within a PRB. This is why we adopt zero-forcing DBF at the BS to cancel the interference between the selected UEs within a PRB as in \cite{UL1_LCMR}, thereby enabling the convexification of the optimization problem to obtain an optimal solution.

Iterative solutions are employed in \cite{UL1_LCMR,UL2_LCHB,UL5_HPCD,UL7_JUAH}, which can lead to high runtime, making them impractical for real-time operation. 

For completeness, we note that \cite{DL} investigates the RRM for the DL of a multi-channel HBF system based on predefined codebooks with $K<U$, but the solutions cannot be reused because RRM on the DL and the UL are inherently different due to power management.

\section{System Model} 

\label{sec:sysmodel}
This section introduces the HBF system based on predefined codebooks that we study, along with a short description of the two operational procedures which supply the necessary inputs for the RRM, namely beam alignment (BA) and channel estimation (CE). 

\begin{table}[!t]
\renewcommand{\arraystretch}{1.3}
\caption{Table of Notation (No.: number, Eff.: effective)} 
\label{Notation}
\centering
\begin{tabular}{|c|c|}
\hline
$N_b$/$N_u$ & No. of antennas at the BS/UE\\
\hline
$K$ & No. of RF chains at the BS\\
\hline
$M$ & No. of matched BS beams per UE during beam alignment\\
\hline
$B_b$/$B_u$ & No. of predefined beams in the BS/UE codebook\\
\hline
$\mathcal{C}_b$/$\mathcal{C}_u$ & BS/UE codebook\\
\hline
$\mathbf{w}_j$/$\mathbf{v}_j$ & the $j$-th beam in BS/UE codebook\\
\hline
$\mathbf{v}_{l_u}$ & selected UE beam of UE $u$ from its codebook\\
\hline
$\mathcal{I}_{s}$ & set of all matched BS beam indices across all UEs\\
\hline
$C$ & No. of channels in a time slot\\
\hline
$Q$ & No. of blocks for CSI reporting in a time slot\\
\hline
$N_F$ & No. of channels in a block ($C/Q$)\\
\hline
$g_{q,n,u,m}^\text{eff}$ & Eff. channel from UE $u$ to $n$ at block $q$ with $\mathbf{w}_{h(n,m)}$\\
\hline
$\mathbf{g}^\text{eff}_{q,u}$ & measured CSI of UE $u$ at block during CE\\
\hline
\end{tabular}
\end{table}

\subsection{The System}

We consider an OFDMA system over a bandwidth made of $C$ channels (a channel and a time slot define a PRB). We will study uplink RRM over a time horizon of $T$ time slots which has to be large enough to take into account channel variations and fairness as will be discussed later. The system is a fully connected HBF~\cite{HBFSurvey}. Specifically, as illustrated by Fig.~\ref{fHBF}, the BS is equipped with $N_b$ antennas, $K$ RF chains and $N_b \times K$ phase shifters. There are $U$ UEs in the cell. Each UE has $N_u$ antennas, one RF chain and $N_u$ phase shifters. We assume $K \ll N_b$ and do not require $K>U$. Beamforming at the BS includes ABF by the phase shifters and DBF by the digital processor. Beamforming at the UE is purely analog because there is only one RF chain. The ABF at the BS is performed by selecting up to one beam for each RF chain from a predefined codebook. Each selected beam specifies the phase shifts between the corresponding RF chain and all BS antennas. 
The BS and UE  codebooks are predefined prior to system operation to reduce real-time computation during RRM \cite{Codebook1Plus,Codebook2Plus}. Specifically, the BS codebook $\mathcal{C}_b$ and UE codebook $\mathcal{C}_u$ are defined as $\mathcal{C}_b=\{\mathbf{w}_j \in \mathbb{C}^{N_b \times 1}:\lVert \mathbf{w}_j \lVert ^2 =1,j=1,...,B_b\}$ and $\mathcal{C}_u=\{\mathbf{v}_j(u) \in \mathbb{C}^{N_u \times 1}:\lVert \mathbf{v}_j(u) \lVert ^2 =1,j=1,...,B_u\}$ respectively. $B_b$ and $B_u$ are the ABF codebook sizes of BS and UE, respectively. In the following, we assume that all UEs use the same codebook and remove the index $u$. For brevity, we also call $\mathbf{w}_j$, BS beam $j$, and $\mathbf{v}_j$, UE beam $j$.

\begin{figure}
    \centering
    \includegraphics[width=70mm]{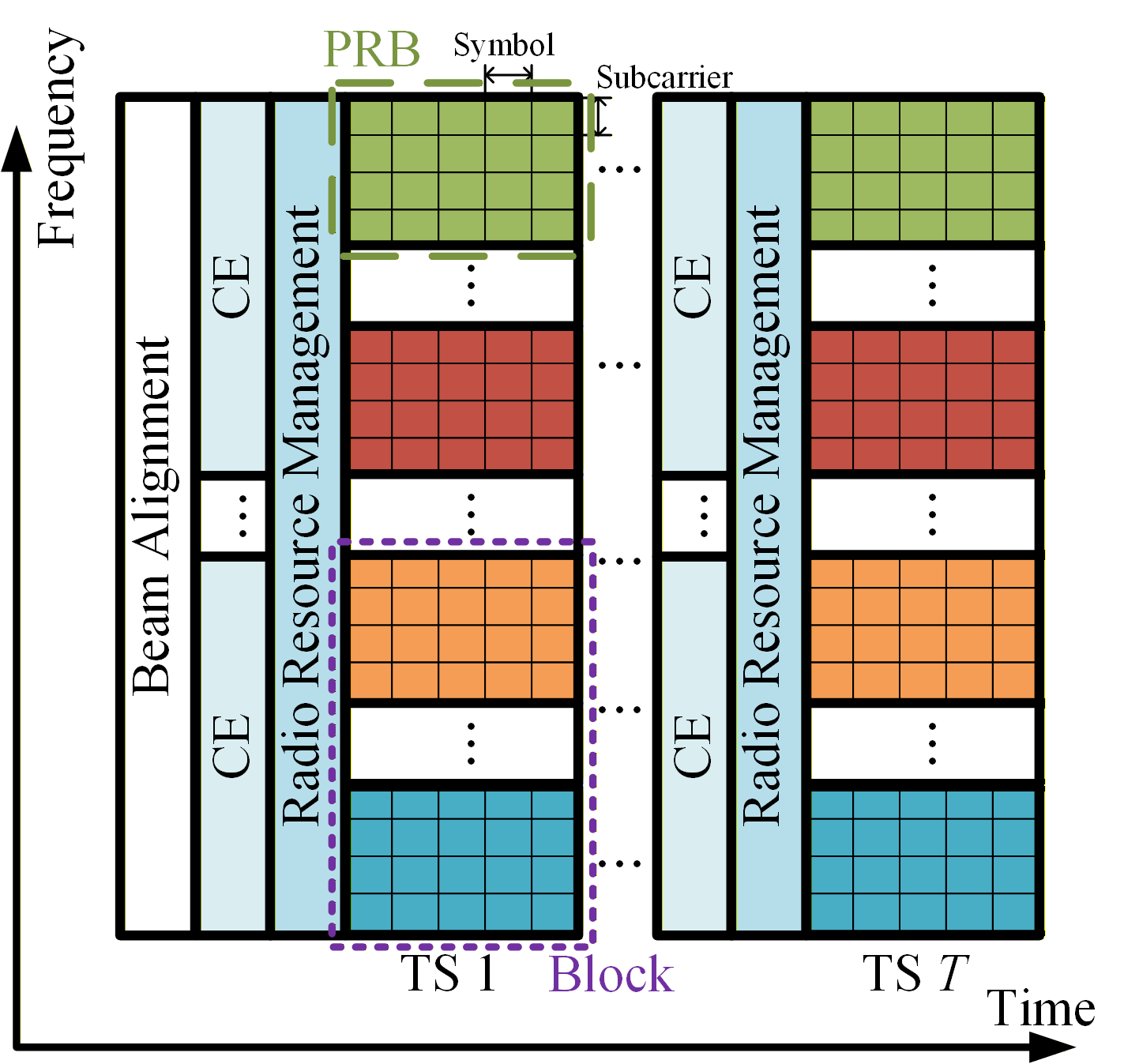} 
    \caption{Data structure and operational procedures.}
    \label{procedure}
\end{figure}

\subsection{Generating Inputs to RRM}
In addition to the fixed system parameters defined above, the uplink RRM process, performed per time slot, relies on three sets of inputs that are updated every time slot or less frequently, as detailed below. 

\subsubsection{First Input Set: Beam Alignment}\label{BA}

The first input set,  determined during beam alignment (BA), includes, for each UE, one matched UE beam along with $M$ matched BS beams. For each UE,  channel estimation (CE) will be performed on the $M$ pairs containing the same UE beam and one of the $M$ matched BS beams. Please note that most papers assume that $M=1$ \cite{UL1_LCMR,UL7_JUAH,ECSIAhmed}. Considering $M>1$ is important for two practical cases. First, when multipaths arrive at the BS with different angles of arrival, which can be captured by different beams. In this case, even a second-best matched BS beam can still offer a sufficiently strong effective channel gain. Second, when a UE lies near the boundary of two beams, making it reasonable for more than one beams beam to be listed as candidates. These two cases are more likely to arise when the BS codebook contains a large number of beams (which is our assumption), and in such scenarios, having $M>1$ can provide meaningful performance improvement.

BA can be realized by various techniques (e.g., \cite{ECSIAhmed,BA3,BA6}) and many criteria can be used to select the matched beams for a UE. The RRM scheme that we will describe later in the paper is agnostic to the BA method. 
Specifically, during BA,  UE $u$ selects 
the pair of beams $(v_{l_u},w_{b_u})$ (where $v_{l_u} \in \mathcal{C}_u$ and $w_{b_u} \in \mathcal{C}_b$) that  offers the highest criterion.  After that, $v_{l_u}$ is fixed and $u$ selects the $M$ BS beams from $\mathcal{C}_b$, which provide the top $M$ highest criteria along with $l_u$. 
We say that these $M$ BS beams are matched to UE $u$. We introduce the mapping $h(u,m)=b$ from UE $u$ to indicate that BS beam $w_b$ is the one that offers the $m$-th highest criterion to UE $u$ with $v_{l_u}$. Accordingly, the set of $M$ matched BS beams for UE $u$ is given by $\{\mathbf{w}_{h(u,1)},...,\mathbf{w}_{h(u,m)},...,\mathbf{w}_{h(u,M)}\}$. Furthermore, we define $\mathcal{I}_s$ as the set of all matched BS beam indices across all UEs, that is, the set of $h(u,m)$  $\forall u \in \mathcal{U}$ and $1 \leq m \leq M$, sorted in ascending order with duplicates removed. We also define $\mathcal{U}(b)$ as the set of UEs that are matched with BS beam $b$, as well as the function $g(u,b)=m \text{ or } 0$ indicating the rank of BS beam $b$ among the BS beams matched with UE $u$ (0 if beam $b$ is not matched with UE $u$ or $u=0$).

BA is an expensive process which is done at a time scale larger than the time slot duration used by RRM. In the following, we assume that BA is done once in the time horizon that we consider and hence, this first set remains fixed in our study.

\subsubsection{Second Input Set: Effective Channel Estimation} \label{ECSI}
The second input set comprises the effective channel state information (ECSI), represented as a $U \times M$ matrix per PRB for each UE. These matrices are derived from CE based on the UE’s matched beams. Note that CE is performed frequently to take channel variations into account and 
that the UEs and the BS do not measure full CSI in the input generation processes.  
Effective CE is typically done every $n$ TSs for each user and if the bandwidth is large (i.e., $C$ is large), it might be done per block of channels. In the following, we assume that effective CE is performed $Q \geq 1$ times per TS.

Specifically, UE $u$ transmits pilot signals 
using $\mathbf{v}_{l_u}$,  $1\leq Q \leq C$ times per time slot (i.e., for each block of $N_F=C/Q$ PRBs in a time slot) \cite{ULDLChannel}. The BS measures the pilot signal of UE $u$ in block $q$ with each matched BS beam $\mathbf{w}_{h(u,m)}$ and computes the $g_{q,n,u,m}^\text{eff}$, which represents the effective channel gain between UE $u$ and the BS using $\mathbf{w}_{h(u,m)}$ if $n = u$ or the effective channel interference between UE $n$ and the BS due to the transmission of UE $u$ to the BS using $\mathbf{w}_{h(n,m)}$ if $n \neq u$,  $\forall m \in \{1,...,M\}$. We define the ECSI (the effective CSI measured during CE) of UE $u$ in block $q$ as $\mathbf{g}^\text{eff}_{q,u} = [\mathbf{g}^\text{eff}_{q,u,1},...,\mathbf{g}^\text{eff}_{q,u,m},...,\mathbf{g}^\text{eff}_{q,u,M}]$ where $\mathbf{g}^\text{eff}_{q,u,m}=[g_{q,1,u,m}^\text{eff},...,g_{q,U,u,m}^\text{eff}]^T \in \mathbb{C}^{U \times 1}$. The ECSI will be used for SINR calculation as  explained  in Section~\ref{problem}. The size of the ECSI matrix per PRB per UE increases linearly with $M$. Therefore, it is crucial to evaluate the impact of $M$ on system performance to determine whether the overhead cost is justified. 

\subsubsection{Third Input Set: Weights to Enforce Fairness}
RRM needs another set of inputs to enforce fairness. These inputs will be described in Section~\ref{FN}.

Note that the RRM techniques that we are about to introduce are independent of the BA and CE methods. In this paper, we posit the perfection of CE. The impact of imperfect CSI will be investigated in future work. Fig.~\ref{procedure} presents an example of the system operation in time and frequency, where BA is performed every 
$T$ time slots, and CE is conducted 
$Q$ times for each block of $N_F$ PRBs within a slot, and RRM is followed by the transmission of 
$Q$ data blocks in that slot.

\section{Joint RRM Optimization} \label{sec:optimization}
In this section, we formulate and solve the joint uplink RRM problem that includes beam selection, user selection, PA, DBF and MCS selection, with an objective function based on proportional fairness. This problem is a large-scale mixed-integer nonlinear program. Using relaxations and approximations, we reformulate it as a linear programming (LP) problem, from which we obtain an upper bound and a feasible solution to the original problem. We will show in Section~\ref{sec:results}, that this feasible solution is quasi-optimal by comparing it with the upper bound. It will serve as the target performance for the offline heuristic RRM scheme that we propose in Section~\ref{sec:heu}. 

The problem formulation and solution approach presented in this work are fundamentally different from those developed in \cite{DL} for the downlink.  First, the power problem on the downlink is a per PRB power distribution problem for the selected streams, while the PA problem that distributes the BS power budget over all PRBs in a time slot is straightforward (each PRB receives the same share of the budget). On the uplink, there is no power distribution (since there is at most one stream per UE) and the PA is challenging. 
Second, to take into account the fact that more than one BS beams are matched with each UE, we need a different representation of  beam and user selections. Specifically, we use  a compact vector notation,  where the position of a user in the user vector indicates for which beam this user is selected in a given PRB. Finally, because of the above differences, a new solution framework had to be developed, in which the RRM problem is formulated and solved jointly as a single integrated optimization problem, rather than decomposed  into independent subproblems.

\subsection{Fairness} \label{FN}
Fairness plays a critical role in RRM research \cite{DL,PF1,PF2,ULFN}. In this paper, we consider proportional fairness (PF) as it is the practice. The RRM is carried out per time slot. At the beginning of time slot $t$, let $R_u(t)$ be the per time slot average throughput received by $u$ over the past $W$ slots. The PF objective function in that time slot can be written as the following weighted sum rate (WSR) maximization \cite{DL,ULFN}
\begin{align}\label{OF}
    \max \sum_{u\in \mathcal{U}} \frac{\lambda_u(t)}{WR_u(t)}, 
\end{align}
where $\lambda_u(t)$ is the rate of UE $u$ in time slot $t$ (i.e., the sum of the rates seen in each PRB of the slot) and the weight of user $u$ is $\frac{1}{WR_u(t)}$. At the end of time slot $t$, we use a moving window of size $W$ to update $R_u(t)$ as $$R_u(t+1)=(1-1/W) R_u(t)+\lambda_u(t)/W, \;\; \forall u \in \mathcal{U}.$$  For brevity of notation, we will remove the slot index $t$ in the following since all our computations are slot-based. 

As discussed in \cite{DL,ULFN}, the natural performance metric for PF is the geometric mean of the rates, noted $GM$ in the following,  i.e., given a time horizon of $T$ time slots,
\begin{align}\label{GMGM}
    GM = \left(\prod_{u=1}^U \frac{1}{T} \sum_{t=1}^{T}  \lambda_u(t)\right)^\frac{1}{U}.
\end{align}

\subsection{Problem Formulation} \label{problem}
We now formulate the UL joint RRM optimization problem that includes beam selection, user selection, PA, DBF and MCS selection.
The problem is formulated for a given time slot, and the inputs are: $K$, $C$, $N_u$, $N_b$, $M$, $(R_u)$, $\mathcal{C}_b$, $\mathcal{C}_u$, $(g(u,b))$, $(h(u,m))$, $\mathcal{I}_s$, and $(\mathbf{g}^\text{eff}_{q,u})$. 

Recall that beam selection consists in selecting $L=\min(K,|\mathcal{I}_s|)$ BS beams for each time slot (the selected beams remain unchanged in all PRBs of the time slot). We define
\begin{align}
    \mathcal{L}=\big\{\ell:\ell=[b_1,..,b_j,..,b_{L}],\ b_j \in \mathcal{I}_s, j \in \{1,...,L\}\big\}
\end{align}
as the  set  of all the possible beam vectors $\ell$ of $L$ selected beams (where $b_j$ is a beam index). Beams in $\ell$ are ranked in an ascending order of the beam indices for convenience. A beam index can appear at most once in a beam vector. For example, if $L=4$ and $|\mathcal{I}_s|=12$, $\ell=[3,6,8,10]$ is a possible beam vector.


Beam selection consists in selecting one beam vector $\ell \in \mathcal{L}$ in the time slot. Then, user selection consists in selecting a user vector  per PRB, given $\ell$. Let the set of all distinct possible user vectors, given $\ell$, be:
\begin{align}
   & \mathcal{M}_\ell =  \big\{z_\ell:z_\ell=[u_1,...,u_j,...,u_L], u_j \in \{\mathcal{U}(b_j),0\}, \nonumber\\
    & \text{ s.t. } u_i = u_k  => u_i = 0, \forall i \neq k  \in \{1,...,L\}  \big\}, \label{userset}
\end{align}
where $z_\ell$ is a user vector in which $u_j$ is the UE selected in the beam in position $j$ in the beam vector $\ell$ (to be selected, $u_j$ needs to be matched to that beam). If $u_j$ is zero, no UE is selected for beam $b_j$. Recall that a UE can  be selected at most once in $z_\ell$.
 
Next, we introduce the beam selection binary variables $\alpha_\ell$ indicating whether beam vector $\ell$ is selected in a given time slot. We have 
\begin{align}
&\sum_{\ell \in \mathcal{L}} \alpha_\ell =1, 
 \label{C1}\\
&\alpha_\ell \in \{0,1\},~
\forall \ell\in\mathcal{L}. \label{Z1}
\end{align}

For each beam vector $\ell$, we define next the user selection variables. Specifically, let $\beta_\ell^{q}(z_\ell)$  represents the fraction of PRBs  within block $q$ allocated to user vector $z_\ell$, given that beam vector $\ell$ has been selected. Note that knowing the exact PRBs occupied by $z_{\ell}$ in a block is inconsequential since the channel coefficients remain the same in a block. We have 
\begin{align}
    & \sum_{z_\ell \in \mathcal{M}_\ell} \beta^q_\ell (z_\ell) \leq 1, \;\; \beta^q_\ell (z_\ell) \geq 0,  ~\forall \ell \in \mathcal{L}, ~q \in \mathcal{Q}, \label{C2}\\
    & N_F \beta^q_\ell(z_\ell) \in \mathbb{Z}, ~\forall \ell \in \mathcal{L}, ~q \in \mathcal{Q},  ~z_\ell \in \mathcal{M}_\ell, \label{Z2}
\end{align}
where $\mathcal{Q} = \{1,2,,\ldots,Q\}$. Eq.~(\ref{Z2}) indicates that each user vector must be allocated an integer number of PRBs. 


With $(\alpha_\ell)$ and $(\beta^q_\ell(z_\ell))$, beam selection and user selection are done. Next is  PA. Let us introduce the PA variables, which determine how each UE’s power budget, $P_\text{UE}$, is allocated across the PRBs on which the UE has been selected for transmission.  Let $p_\ell^{q,u} (z_\ell)$ be the power allocated to each PRB in block $q$ on which $z_\ell$ is selected. We have  
\begin{align}
    & 0 \leq p_\ell^{q,u} (z_\ell) \leq P_\text{UE}, \forall u \in z_\ell, \ell, z_\ell, q,  \label{ULPower1}\\
    & p_\ell^{q,u} (z_\ell) = 0, \forall u \notin z_\ell, u=0, \ell, z_\ell, q. \label{ULPower0}
\end{align}

Unlike the DL where there is a per PRB power constraint linking  all selected UEs in the PRB \cite{DL}, on the uplink, the power constraint is per UE and per time slot, i.e.:
\begin{align}
    & \sum_{q \in \mathcal{Q}} \sum\limits_{z_\ell \in \mathcal{M}_\ell } N_F \beta^q_\ell (z_\ell) p_\ell^{q,u}(z_\ell) \leq P_\text{UE}, ~\forall u, \ell, \label{ULPower2}
\end{align}
where $N_F \beta^q_\ell (z_\ell)$ is the number of PRBs allocated to vector $z_\ell$ in block $q$.

We present next the digital BF step, i.e., ZF-DBF. Note that ZF-DBF and the corresponding ECSI are precalculated for all possible $z_\ell$ and $\ell$. In other words, given $z_\ell$ and $\ell$, the effective channel after ZF-DBF and ABF is computed, and ZF-DBF is therefore implicitly embedded in the joint RRM formulation.
Specifically, to calculate the ECSI after  ZF-DBF, we first remove the zero element(s) in $z_\ell$ and the corresponding beam(s) in $\ell$ to get $\ell'=[b_1,...,b_k,...,b_{L'}]$ and $z_{\ell'}=[u_1,...,u_k,...,u_{L'}]$. The new ECSI vector of UE $u_k$ is then:
\begin{align}
    \mathbf{g}_{\ell'}^{q,u_k}(z_{\ell'})=[g_{q,u_1,u_k,m_{u_k}}^\text{eff},...,g_{q,u_{L'},u_k,m_{u_k}}^\text{eff}], \nonumber
\end{align}
where $m_{u_k}=g(u_k,b_k)$. Then the ECSI matrix for user vector $z_{\ell'}$ is defined as
\begin{align}
    \mathbf{G}_{\ell'}^q(z_{\ell'})=[\mathbf{g}_{\ell'}^{q,u_1}(z_{\ell'}),...,\mathbf{g}_{\ell'}^{q,u_{L'}}(z_{\ell'})]^T. \nonumber
\end{align}
We also have the ABF matrix
\begin{align}
    \mathbf{F}_{\ell'}^{\text{ABF}}(z_{\ell'})=\big[\mathbf{w}_{h(u_1,m_{u_1})},...,\mathbf{w}_{h(u_k,m_{u_k})},...,\mathbf{w}_{h(u_{L'},m_{u_{L'}})}\big]. \nonumber
\end{align}
We define the ZF-DBF vector of UE $u_k \in z_{\ell'}$ as 
\begin{align}
    \mathbf{d}_{\ell'}^{q,u_k}(z_{\ell'})=\frac{\mathbf{a}_{\ell'}^{q,u_k}(z_{\ell'})}{\lVert \mathbf{F}_{\ell'}^{\text{ABF}}(z_{\ell'}) \mathbf{a}_{\ell'}^{q,u_k}(z_{\ell'}) \lVert_F} \nonumber
\end{align}
where $\mathbf{a}_{\ell'}^{q,u_k}(z_{\ell'})$ is calculated from
\begin{align}
    & [\mathbf{a}_{\ell'}^{q,u_1}(z_{\ell'}),...,\mathbf{a}_{\ell'}^{q,u_k}(z_{\ell'}),...,\mathbf{a}_{\ell'}^{q,u_{L'}}(z_{\ell'})] \nonumber \\
    & =\mathbf{G}_{\ell'}^q(z_{\ell'})^H [{\mathbf{G}_{\ell'}^q(z_{\ell'})}\mathbf{G}_{\ell'}^q(z_{\ell'})^H]^{-1}. \nonumber
\end{align}
Then the ECSI between UE 
$u_k$ and $u_j$ in user vector $z_{\ell'}$ in  block $q$ after applying both ABF and DBF is
\begin{align}
h_{\ell'}^{q,u_k,u_j}(z_{\ell'})=\mathbf{g}_{\ell'}^{q,u_j}(z_{\ell'})^T \mathbf{d}_{\ell'}^{q,u_k}(z_{\ell'}), \nonumber
\end{align}
which denotes the interference of UE $u_k$ caused by UE $u_j$ transmitting to the BS or the signal of UE $u_k$ if $u_j=u_k$. Combined with zero element(s) in $z_{\ell}$, we have
\begin{align}
h_\ell^{q,u_k,u_j}(z_{\ell})=\left\{
    \begin{array}{l}
    h_{\ell'}^{q,u_k,u_j}(z_{\ell'}), \text{if } u_j,u_k \neq 0\\
    0, \text{otherwise }\\
    \end{array}
    \right.. \nonumber
\end{align}
We assume $h_\ell^{q,u_k,u_j}(z_{\ell}) = 0$ for $u_k \neq u_j$ as a result of zero-forcing digital beamforming (ZF-DBF). Note that this assumption is not always true, particularly when the effective channel matrix of the selected UEs is not full rank (i.e., when the selected UEs exhibit poor orthogonality). In our study, we adopt the codebook proposed in \cite{CODEBOOK3} to ensure good orthogonality between beams, and consequently between the selected UEs. We do not impose any assumption on the rank of the channel matrix. Extensive numerical results confirm that the residual inter-user interference after applying ZF-DBF is negligible. Consequently, we use SNR instead of signal-to-interference-plus-noise ratio (SINR) when mapping to a UE's rate.

Assuming $\ell$ and $z_\ell$ are selected in a PRB within block $q$, and the allocated power to UE $u$ is $p_{\ell}^{q,u}(z_\ell)$, we calculate the SNR of UE $u$ in this PRB as
\begin{align}
    & \gamma_\ell^{q,u}(z_\ell)=
    \frac{|h_\ell^{q,u,u}(z_\ell)|^2 p_{\ell}^{q,u}(z_\ell)}{\sigma_\text{PRB}^2}, \forall u\label{C7l}
\end{align}
where $\sigma_\text{PRB}^2$ is the noise power on a PRB.
The rate function $f(.)$ maps $\gamma_\ell^{q,u}(z_\ell)$ to a rate. We use a rate function based on practical modulation and coding schemes (MCSs) \cite{DL,ULFN,MCS}  rather than Shannon capacity formula. Specifically, we model the rate function as a piecewise constant function of SNR, consisting of $L$ levels corresponding to the $L$ MCSs, i.e.,
\begin{equation} f(\gamma)=B_c\big(s_1\mathbbm{1}_{[\Gamma_1,\Gamma_2)}(\gamma) + 
  \cdots + s_{L} \mathbbm{1}_{[\Gamma_{L},\infty)}(\gamma)\big), \label{eq:true_rate}
\end{equation} 
where $B_c$ is the bandwidth of a PRB  in Hz, $\mathbbm{1}_A(x)$ is the indicator function, which equals 1 if $x \in A$ and zero otherwise, $\Gamma_l$ denotes the SINR decoding threshold for MCS $l$ and $s_l$ is the spectral efficiency of MCS $l$ measured in bps/Hz. Note that our RRM problem formulation and solution are independent of the choice of MCSs and SINR thresholds. For the numerical results, we adopt a widely used MCS-based rate function, as explained later in Section~\ref{sec:results}. 

\color{black}

With $\gamma_\ell^{q,u}(z_\ell)$ and $f(.)$, the UL data rate of UE $u$ in a PRB within block $q$ allocated to UE set $z_\ell$ is defined as
\begin{align}
r_\ell^{q,u}(z_\ell) =  f\big(\gamma_\ell^{q,u}(z_\ell)\big), ~\forall q,u,\ell,z_\ell. \label{ULRate}
\end{align}
Given $\alpha_\ell$, $\beta^q_\ell(z_\ell)$ and $r_\ell^{q,u}(z_\ell)$, we can compute the throughput $\lambda_u$ of UE $u$ in the time slot as
\begin{align}
    \lambda_u = \sum_{\ell \in \mathcal{L}} \alpha_\ell \sum_{q \in \mathcal{Q}} N_F \sum_{z_\ell \in \mathcal{M}_\ell} \beta^q_\ell (z_\ell) r_\ell^{q,u}(z_\ell), \forall u \in \mathcal{U}. \label{ULlambda}
\end{align}
Next, we formulate the general RRM optimization problem, $\mathbf{\Omega}$,  for a time slot. Given $f(.)$, $\mathcal{U}$, $\mathcal{Q}$, $\mathcal{L}$, $\mathcal{M}_\ell$, $R_u$, $W$, $N_F$ and $h_\ell^{q,u,u}(z_\ell), \forall z_\ell,\ell,q, u\in z_\ell$, we formulate
\begin{align}
    \mathbf{\Omega}: & \max_{\beta^q_\ell(z_\ell), \alpha_\ell, p_{\ell}^{q,u}(z_\ell), \gamma_\ell^{q,u}(z_\ell), r_\ell^{q,u}(z_\ell), \lambda_u} \sum_{u\in \mathcal{U}} \frac{\lambda_u}{W R_u}, \nonumber \label{}\\
    ~~\text{s.t. } & (\ref{C1}), (\ref{Z1}), (\ref{C2}), (\ref{Z2}), (\ref{ULPower1}), (\ref{ULPower0}), (\ref{ULPower2}), (\ref{C7l}), (\ref{ULRate}), (\ref{ULlambda}) \nonumber
\end{align}
Problem $\mathbf{\Omega}$ is non-convex, non-linear and includes integer constraints.

\subsection{Upper Bound of Joint RRM Optimization Problem} \label{UpperBound}

In order to provide an upper bound on the original problem $\mathbf{\Omega}$, we first introduce $I$ linear functions, following the method in \cite{MCS1}, to upper-bound the nonconvex rate function $f(\gamma)$ as 
\begin{align}
    f(\gamma) \leq a_i\gamma+b_i \;\;\; \forall i \in \{1,2,...,I\}. \label{approx_rate}
\end{align}
Note that $a_1>...>a_I \geq 0$ and $b_I>..>b_1 \geq 0$. By substituting approximation~(\ref{approx_rate}) in Eq.~(\ref{ULRate}), we have
\begin{align}
r_\ell^{q,u}(z_\ell) \leq a_i \gamma_\ell^{q,u}(z_\ell) + b_i, \forall i,  q,u,\ell,z_\ell. \label{ULRateA}
\end{align}
Using this linear upper bound, rather than the logarithmic upper bound in \cite{DL}, is essential for enabling the subsequent variable transformations and to convexify the problem into a LP problem, as discussed later in this section.

We also remove the integer constraints \eqref{Z1} and \eqref{Z2} in $\mathbf{\Omega}$ to obtain the following relaxed problem
$\mathbf{\Omega}^\text{R}$: 
\begin{subequations}
\begin{align}
    &\mathbf{\Omega}^\text{R}:  \max_{\beta^q_\ell(z_\ell), \alpha_\ell, p_{\ell}^{q,u}(z_\ell),  r_\ell^{q,u}(z_\ell), \lambda_u} \sum_{u\in \mathcal{U}} \frac{\lambda_u}{W R_u}, \nonumber \\
    & \sum_{\ell \in \mathcal{L}} \alpha_\ell =1, 0 \leq \alpha_\ell \leq 1, ~\forall \ell \in \mathcal{L}, 
    \\
    & \sum_{z_\ell \in \mathcal{M}_\ell} \beta^q_\ell (z_\ell) \leq 1, \;\; \beta^q_\ell (z_\ell) \geq 0,  ~\forall \ell \in \mathcal{L}, ~q \in \mathcal{Q}, \label{C2v}\\
    & 0 \leq p_\ell^{q,u} (z_\ell) \leq P_\text{UE}, \forall u \in z_\ell, \ell, z_\ell, q, \label{ULPower1v} \\ 
    & p_\ell^{q,u} (z_\ell) = 0, \forall u \notin z_\ell, u=0, \ell, z_\ell, q,  \label{ULPower0v} \\ 
    & \sum_{q \in \mathcal{Q}} \sum\limits_{z_\ell \in \mathcal{M}_\ell } N_F \beta^q_\ell (z_\ell) p_\ell^{q,u}(z_\ell) \leq P_\text{UE}, ~\forall u, \ell, \label{ULPower2v}\\ 
    & \lambda_u = \sum_{\ell \in \mathcal{L}} \alpha_\ell \sum_{q \in \mathcal{Q}} N_F \sum_{z_\ell \in \mathcal{M}_\ell} \beta^q_\ell (z_\ell) r_\ell^{q,u}(z_\ell), \forall u \in \mathcal{U}, \label{ULlambdav}\\
    & r_\ell^{q,u}(z_\ell) \leq  a_i
    \frac{|h_\ell^{q,u,u}(z_\ell)|^2 p_\ell^{q,u}(z_\ell)}{\sigma_\text{PRB}^2}  +b_i , \forall i, u , \ell, q, z_\ell. \label{ULRateUpper}
\end{align}
\end{subequations}
Please note that $\mathbf{\Omega}^\text{R}$ remains highly nonconvex due to two products of variables, one between the user selection and PA variables in Constraint~\eqref{ULPower2v}, and the other between beam selection, user selection, and PA variables in Constraint~\eqref{ULlambdav}. We introduce the following three changes of variables to convexify the problem:
\begin{align}
    & \eta^{q,u}_\ell (z_\ell) \triangleq \alpha_\ell \beta^q_\ell (z_\ell) p_\ell^{q,u}(z_\ell), & \forall u , \ell, q, z_\ell, \label{n}\\
    & \epsilon^{q,u}_\ell (z_\ell) \triangleq \alpha_\ell \beta^q_\ell (z_\ell) r_\ell^{q,u}(z_\ell), & \forall u , \ell, q, z_\ell, \label{e}\\
    & \xi^q_\ell (z_\ell) \triangleq \alpha_\ell \beta^q_\ell (z_\ell), & \forall u , \ell, q, z_\ell. \label{E}
\end{align}
The problem then becomes
\begin{subequations}
\begin{align}
    &\mathbf{\Omega}^\text{R}:  \max_{ \eta^{q,u}_\ell (z_\ell), \epsilon^{q,u}_\ell (z_\ell), \xi^q_\ell (z_\ell), \alpha_\ell, \lambda_u} \sum_{u\in \mathcal{U}} \frac{\lambda_u}{W R_u}, \nonumber \\
    & \sum_{\ell \in \mathcal{L}} \alpha_\ell =1, 0 \leq \alpha_\ell \leq 1, ~\forall \ell \in \mathcal{L},  \label{C1vv}\\
    & \sum_{z_\ell \in \mathcal{M}_\ell} \xi^q_\ell (z_\ell) \leq \alpha_\ell, \;\; 0 \leq \xi^q_\ell (z_\ell) \leq 1,  ~\forall \ell \in \mathcal{L}, ~q \in \mathcal{Q}, \label{C2vv}\\
   & \sum_{q \in \mathcal{Q}} \sum\limits_{\begin{subarray}{c} z_\ell \in \mathcal{M}_\ell \\ \text{ s.t. } u \in z_\ell \end{subarray}} N_F \eta^{q,u}_\ell (z_\ell) \leq \alpha_\ell P_\text{UE}, \forall u \in \mathcal{U}, \ell \in \mathcal{L}, \label{ULPower2bvv} \\
    & 0 \leq \eta_{\ell}^{q,u}(z_\ell) \leq P_\text{UE} \xi^q_\ell(z_\ell), \forall u \in z_\ell, \ell, z_\ell, q, \label{ULPower1b}\\
    & \eta_{\ell}^{q,u}(z_\ell) = 0, \forall u \notin z_\ell, u=0, \ell, z_\ell, q \label{ULPower0bvv}\\
    & \lambda_u = N_F\sum_{\ell \in \mathcal{L}}  \sum_{q \in \mathcal{Q}}  \sum_{z_\ell \in \mathcal{M}_\ell} \epsilon^{q,u}_\ell (z_\ell), \forall u \in \mathcal{U}, \label{ULlambdavv}\\
    & \epsilon^{q,u}_\ell (z_\ell) \leq a_i \frac{|h_\ell^{q,u,u}(z_\ell)|^2 \eta_{\ell}^{q,u}(z_\ell)}{\sigma_\text{PRB}^2}+ b_i \xi^q_\ell (z_\ell), \forall u, \ell, q, z_\ell, i. 
\end{align}
\end{subequations}

Problem $\mathbf{\Omega}^\text{R}$ then becomes a linear problem (LP) that can be efficiently solved  with commercial solvers such as CPLEX, MINOS and Gurobi \cite{AMPL,Solver}.  Solving it will give us an upper bound on the original Problem $\mathbf{\Omega}$. However, due to the relaxations applied to the rate function and the integer constraints, this upper bound might not be directly feasible. In the next subsection, we present a method to construct a feasible solution using the values of $\alpha_\ell$ and $\xi^q_\ell(z_\ell)$ obtained from solving $\mathbf{\Omega}^\text{R}$.

\subsection{Feasible Solution to Joint RRM Optimization Problem} \label{FS}

\subsubsection{Reconstructing $\alpha_\ell$ and $\beta^q_\ell(z_\ell)$} \label{ReCons}

We first restore the change of variables by $\beta^q_\ell(z_\ell) = \xi^q_\ell(z_\ell)/\alpha_\ell$ if $\alpha_\ell \neq 0$ and $\beta^q_\ell(z_\ell)=0$ if $\alpha_\ell=0$. However, the ($\alpha_\ell$)'s and ($\beta^q_\ell (z_\ell)$)'s do not necessarily satisfy Constraints~\eqref{Z1} and~\eqref{Z2} which were relaxed to get to Problem~$\mathbf{\Omega}^{\text{R}}$. Therefore,  we modify the solution to meet the constraints. Specifically, we first set $\hat{\alpha}_\ell = 1$ for the beam vector $\ell_f$ corresponding to the highest value of $\alpha_\ell$, and set $\hat{\alpha}_\ell = 0$ for all other $\ell$. We then generate $\hat{\beta}_{\ell_f}^q(z_{\ell_f}$) by rounding down $\beta^q_{\ell_f} (z_{\ell_f})$ to satisfy the integer constraint. Then, we choose the $z_{\ell_f} \in \mathcal{M}_{\ell_f}$ such that the gap between $\beta_{\ell_f}^q(z_{\ell_f})$ and $\hat{\beta}_{\ell_f}^q(z_{\ell_f})$ is the largest, and increase $\hat{\beta}_{\ell_f}^q(z_{\ell_f})$ by $1/N_F$. We repeat the last step until $\sum_{z_{\ell_f} \in \mathcal{M}_{\ell_f}} \hat{\beta}_{\ell_f}^q(z_{\ell_f}) =1$. $\hat{\alpha}_{\ell_f}$ and $\hat{\beta}_{\ell_f}^q(z_{\ell_f})$ are the reconstructed beam selection and user selection variables which satisfy the integral constraints.

\subsubsection{Computing a Feasible Power Allocation} \label{RePA}
Note that the previous PA solution $p_{\ell_f}^{q,u}(z_{\ell_f})$ was calculated based on $\alpha_{\ell_f}$ and $\beta_{\ell_f}^q(z_{\ell_f})$ and thus is no longer correct. Therefore, we solve Problem $\mathbf{\Omega}^\text{R}_\text{LP}$ again given $\hat{\alpha}_{\ell_f}$ and $\hat{\beta}_{\ell_f}^q(z_{\ell_f})$ to get a feasible PA solution $\bar{p}_{{\ell}_f}^{q,u}(z_{{\ell}_f})$. After that, we use the MCS-based greedy algorithm proposed in \cite{MCS} to improve the solution by taking the real MCS function into account.  

We have now shown how to compute $\hat{\alpha}_{{\ell}_f}$, $\hat{\beta}_{{{\ell}_f}}^q(z_{{\ell}_f})$ and $\hat{p}_{{\ell}_f}^{q,u}(z_{{\ell}_f})$, which, together, form a feasible solution to the original Problem~$\mathbf{\Omega}$. We will validate the quasi-optimality of this solution by comparing numerically its performance to the upper bound in Section~\ref{sec:results}.
However, solving Problem~$\mathbf{\Omega}^\text{R}$ becomes impossible when the number of users, $U$, is larger than 10 (especially when the number of RF chains is large) due to the large number of variables. For instance, with 20 UEs, there can be $2^{20}$ user selection variables and $2^{20}$ PA variables per block. However, to provide meaningful engineering insights, we need to study systems with a larger number of users than 10 and hence, we propose an offline heuristic, in the next section, to address the computation of the performance for these systems. We will validate it by comparing its solutions to the quasi-optimal solutions for $U \leq 10$ and then provide engineering insights using it. 
We end the next section with a proposal for an online heuristic RRM solution obtained by simplifying the offline version. In Section~\ref{sec:results}, we will present our numerical results in a mmWave setting and show that the online heuristic has a low complexity suitable for real-time system operation, while achieving performance comparable to the offline heuristic.

\section{Heuristic Solutions to RRM} \label{sec:heu}
In this section, we first present our offline heuristic solution, developed to deliver target performance and engineering insights in scenarios with a large number of UEs, where solving the optimization problem $\Omega^\text{R}$ becomes computationally infeasible due to the explosion in the number of variables.

\subsection{Offline Heuristic}

We have observed that simple approaches, such as round-robin beam and user selections combined with equal PA where each UE allocates equal power to its allocated PRBs, do not approach the target performance obtained from optimization. Thus, we propose an offline heuristic that achieves a performance close to the target and can be executed efficiently even with a large number of UEs.


Our offline heuristic is based on four steps, the two main steps being Step~1 and Step~3.  Step~1  optimizes user selection/PA performed per beam \emph{in isolation} (while ignoring the other beams), considering all users that have matched with this beam. Step~3 is a ``load-aware'' beam selection strategy where the $K$ beams with the largest loads are selected.

\subsubsection{Step~1: Per-beam User Selection and Power Allocation Optimization}
For every $b \in \mathcal{I}_s$ and the set $\mathcal{U}(b)$ of users that are matched to $b$, we schedule users in a PF fashion, ignoring the other beams and hence the need for DBF. The per time slot scheduling problem can be written as problem $\mathbf{S_b}$, where we have removed the index for $b$ on all variables except in $\mathcal{U}(b)$ for ease of notation.
\begin{subequations}
\begin{align}
   &\mathbf{S_b}: \max_{m^q_u,  p_{u}^{q},  r_u^{q}, \lambda_u} \sum_{u\in \mathcal{U}(b)} \sum_{q \in \mathcal{Q}}   \frac{m^q_u  r_u^{q}}{R_u}, \nonumber \\
    & \sum_{u\in \mathcal{U}(b)}  m^q_u \leq N_F,~~\forall q \in \mathcal{Q} \\
    & m_{u}^{q} \in \mathbb{Z} , ~~\forall u \in \mathcal{U}(b), q \in \mathcal{Q} \\
    & 0 \leq p_{u}^{q} \leq P_\text{UE}, ~~\forall u \in \mathcal{U}(b), q \in \mathcal{Q}\\ 
     & \sum_{q \in \mathcal{Q}} m^q_u p^q_u \leq P_\text{UE}, ~~\forall u \in \mathcal{U}(b) \\
    & r_u^{q} \leq  a_i
    \frac{G_u^q ~p_{u}^{q}}{\sigma_\text{PRB}^2}  +b_i , ~~\forall i, u ,  q. 
\end{align}
\end{subequations}
where $m_u^q$ is the number of PRBs in block $q$ allocated to UE $u$, $p_{u}^{q}$ is the power that UE $u$ allocates to a PRB in block $q$, and $G_u^q= g^\text{eff}_{q,u,u,g(u,b)}$ is the channel gain between UE $u$ and BS beam $b$. 
This problem is a simplified version of Problem~$\Omega^\text{R}$ and hence can  be relaxed and transformed as discussed earlier for $\Omega^\text{R}$ to yield an LP problem that is small in size since the number of users matched to a beam is usually small (except for $M$ large).
\color{black}


The outputs of this per-beam optimization are a selected UE (or no UE) per PRB, drawn from the UEs matched with the beam, the power allocated to a UE in its allocated PRBs, and a weighted sum rate per beam. We use this weighted sum rate as the ``load'' of beam $b$. 

\subsubsection{Step~2: Only necessary if $M>1$} Note that, if $M>1$, a UE might be selected multiple times in a PRB (on different beams). In that case, we say that the UE is in a conflict. These conflicts have to be removed to provide a feasible solution over all beams. If a UE is in a conflict in  a given PRB, we keep it on the beam providing the highest channel gain. Then, we  try to assign different PRBs in the same block to the UE in the beams in conflict. For example, if UE $a$ has been  removed from beam $i$ in the PRB. We first check if there is any empty PRB in beam $i$ in this block, which is not in conflict with UE $a$. If yes, we move the UE to this PRB in beam $i$. If no, we check each of the other PRBs in the block in beam $i$ one by one. As soon as we find one that is occupied by say UE $b$ such that switching UE $a$ and UE $b$ gives feasible user sets in both PRBs (i.e., no new conflicts are created),  we switch $a$ and $b$. If there is no such $b$, we remove UE $a$.  This ensures that each UE is selected at most once in a PRB and prevents duplicate counting of a UE's contribution to the weighted sum rate calculations across beams.

Once we have removed all conflicts, we need to recompute the PA for each beam for which there were removed users to obtain the revised loads for these beams.

\subsubsection{Step~3: Load-aware Beam Selection}
We select the $L$ beams with the highest weighted sum rates (load) as the beam set vector for the time slot. Together with the selected UE per PRB from per-beam optimization for each of the $L$ beams, we build the user set vectors for all PRBs in the time slot. Recall that the beam set and user set are vectors because beams are one-to-one mapped to UEs in order.  


 

\subsubsection{Step~4: DBF and Final Power Allocation}
With the selected beam set vector and the corresponding per-PRB user set vectors, we apply ZF-DBF  per PRB on the selected UEs to mitigate the inter-beam interference. Since the PA calculated in Step~1 is based on effective channel gains without ZF-DBF, we need to update the PA for each UE. We follow Section~\ref{RePA} for the PA update. Finally, the SNR of each UE is mapped to a rate according to the MCS-based rate function.

In conclusion, we first run per-beam user selection and PA optimization and load-aware beam selection to construct a feasible beam set vector for a time slot and a user set vector for each PRB in the time slot. Next, ZF-DBF is applied per PRB to control the interference between selected UEs. Based on the resulting channel gains after ZF-DBF, per-user PA is updated using the method described in Section~\ref{RePA}. By dividing the joint optimization in Section~\ref{sec:optimization} into separate per-beam optimizations, we significantly reduce the computation time from months to hours to obtain a figure such as Fig.~4.

\subsection{Online Heuristic}
The offline heuristic serves as an effective tool for system planning with a large number of UEs. However, it is not suitable for online deployment since it relies on  per-beam user selection and power optimization, i.e., problem $\mathbf{S_b}$. In order to design an online heuristic, we need to simplify the offline heuristic and, in particular, replace solving $\mathbf{S_b}$ for each beam $b$ with matched users by an online scheduler. Several papers have addressed the problem of uplink scheduling in a single cell but almost none incorporates the practical MCS-based rate function, instead relying on a Shannon-based one. Without considering practical MCS levels, a UE may be selected on too many channels and spread its power thinly among them, resulting in some channels having SINR values too low to achieve a positive rate. Therefore, we first adapt the iterative algorithm developed in \cite{PerBeam} to the MCS-based rate function and see an improvement up to 20\%, but it did not perform well when the number of UEs is small, which is commonly the case when focusing on a single beam. Hence, we propose next, a per-beam scheduler (PBS) that extends the algorithm by incorporating additional search steps considering practical MCS levels.

\subsubsection{Per-beam Scheduler, PBS} \label{PBS}


 The aim of PBS is to select, in a given BS beam in $\mathcal{I}_s$,  at most one UE in each PRB of a time slot from the UEs who are matched with this BS beam. Specifically, PBS is iterative and selects UEs in a greedy way, taking their weights into account. In each step, PBS considers each UE one at a time. It gives the UE under consideration its highest gain PRB among those not yet allocated, and calculates, given all the PRBs allocated to that UE so far,  the weighted rate for that UE in the time slot assuming EPA. Then, PBS finds the UE whose weighted sum rate (sum over its allocated PRBs) increases the most or decreases the least if no UE sees an increase in weighted rate and gives it its preferred PRB. The fact that we allow the increase to be negative is what makes the algorithm perform well when the number of users is small when compared to the original method in \cite{PerBeam}, which terminates immediately upon encountering a negative rate increment. The algorithm stops considering a UE   when the UE has seen $X$ steps with a negative change in the weighted rate. The algorithm also stops once there are no more PRBs to allocated or no UEs to consider.  The improvement arises from the fact that while adding a single PRB may not increase a UE’s rate, adding multiple PRBs can collectively yield a gain, considering the step-wise rate function. Details are presented in Algorithm~\ref{PerBeam}.

\begin{algorithm}
\caption{PBS on beam $b$, Inputs: $\mathcal{U}(b)$, $X$, $R_u$, $P_{\text{UE}}$, $g_{q,u,u,m}^\text{eff}$ s.t. $h(u,m)=b$}
\label{PerBeam}
\begin{algorithmic}[1]
    \State Set the stop flags of all UEs in $\mathcal{U}(b)$ to zero and all PRBs are empty.
    \While{there are any remaining PRBs to allocate in this time slot \textbf{and} UEs whose stop flag is no larger than $X$}
    \For{each UE in $\mathcal{U}(b)$ whose stop flag $\leq X$}
        \State Select the best remaining PRB ($\max_c  g_{q,u,u,m}^\text{eff}$). \label{lp1}
        \State Calculate the sum rate of this UE in this time slot assuming EPA and that this PRB is allocated to it.
        \State Increase its stop flag by one if the sum rate of this UE decreases.
    \EndFor
    \State Select the UE whose weighted sum rate increases the most and allocate its best remaining PRB to it.
    \State Record the user selection of beam $b$ and corresponding weighted rate in each PRB if the weighted sum rate in this time slot is the highest over the past. \label{lp2}
    \EndWhile
 
\end{algorithmic}
\end{algorithm}

To further accelerate Algorithm~\ref{PerBeam}, we can allocate a block of PRBs (recall that there are $Q$ such blocks) to a UE in each step. Once allocating a full block no longer yields performance improvement, we switch to adding one PRB per step. In this case, if we see a sum rate decrease while adding a block of PRBs, we increase the stop flag by $N_F$. In Section~\ref{sec:results}, we will compare the performance of PBS and the algorithm in \cite{PerBeam}.

\subsubsection{The Online Heuristic}
We are now ready to described our online heuristic that mimics the offline one. PBS is used in Step~1, We do not modify Step~2 and Step~3 described above. In Step~4, we replace the last PA by one based on WF \cite{WF}.
Finally, the resulting SNRs are mapped to achievable rates using an MCS-based rate function.

\section{Numerical Evaluations} \label{sec:results}

In this section, we first present numerical results for both the upper bound and the feasible solution of our joint RRM optimization problem to validate the quasi-optimality of our feasible solution for $U \leq 10$. We also analyze the impact of the number of RF chains at the BS ($K$) and the number of matched BS beams per UE during CE ($M$) on system performance based on the feasible solution. Next, we show the effectiveness of our offline RRM heuristic by comparing it with the quasi-optimal feasible solution in scenarios with a small number of UEs. Thanks to its lower complexity, the offline heuristic can be used for scenarios with more UEs, offering engineering insights for practical deployments. Finally, we evaluate the performance and complexity of our online RRM heuristic by comparing it with the offline heuristic. All our numerical results are obtained in a mmWave setting.

\subsection{System Parameters} \label{param}

\begin{table*}
\centering
\caption{$a_i$ and $b_i$ for the linear approximation}
\label{linear}
\begin{tabular}{|c|c|c|c|c|c|c|c|c|c|c|c|c|c|}
\hline
\rowcolor[HTML]{D3D3D3} $i$ & 1 & 2 & 3 & 4 & 5 & 6 & 7 & 8 & 9 & 10 & 11 & 12 & 13 \\
\hline

\cellcolor[HTML]{D3D3D3} $a_i$ & 0.6531 & 0.5652 & 0.4281 & 0.2758 & 0.2387 & 0.1716 & 0.1108 & 0.0752 & 0.0637 & 0.0475 & 0.0279 & 0.0145 & 0 \\
\hline
\cellcolor[HTML]{D3D3D3} $b_i$ & 0 & 0.0658 & 0.2292 & 0.5129 & 0.6223 & 0.9238 & 1.3621 & 1.7499 & 1.9448 & 2.3677 & 3.1290 & 3.9432 & 5.5547 \\
\hline

\end{tabular}
\end{table*}

We consider the uplink of a small cell network with a radius of \( r = 75 \) meters. UEs are uniformly distributed within this area, excluding a 6-meter radius around the BS where standard path loss models are no longer valid. The BS is elevated at a height of 10 meters. Each UE is equipped with a transmit power budget of \( P_{UE} = 7 \) dBm, and the noise power spectral density is set to \( N_0 = -174 \) dBm/Hz. Both the BS and UEs employ uniform linear antenna arrays, comprising $N_b=128$ and $N_u=16$ antennas, respectively, with half-wavelength spacing. The system operates at a carrier frequency of $f_c=28$ GHz with a total bandwidth of $BW=100$~MHz, divided into \( C = 132 \) channels, each with a bandwidth of $B_c=720$ kHz. The time slot duration is 0.25 msec. Our MCS-based rate function consists of 15 MCS levels, each specifying a modulation scheme, code rate, and spectral efficiency, as defined in Table 5.2.2.1-2 of \cite{3GPPMCS}. We adopt the SINR threshold as in \cite{DL,MCS} for each MCS level and use the parameters in Table~\ref{linear} for its approximation~\eqref{approx_rate}.

Analog beamforming (ABF) codebooks for both the BS and UEs are generated following the method in \cite{DL,CODEBOOK1}. We generate 32 beams (codewords) in the BS codebook and 4 beams (codewords) in the UE codebook.

\subsection{Channel Model and Parameters}
We adopt a wideband channel model which incorporates multipath effects, i.e., scattering clusters and spatial paths in each cluster \cite{DL,Channel3}. The channels among blocks of PRBs are correlated because the same multipath components are present across the entire bandwidth, and they differ only due to the frequency of each block. Specifically, the MIMO channel between BS and UE $u$ at block $q$ is defined as the following $N_u \times N_b$ matrix:
\begin{align}
    \mathbf{G}_{q,u} = \frac{1}{\sqrt{N_\text{path}}} \sum_{d=1}^{N_\text{cluster}} \sum_{l=1}^{N_\text{path}} g_{d,l}^{u} e^{-j 2\pi \tau_{d,l}^u f_q} \mathbf{a}_\text{RX}(\phi_{d,l}^u) \mathbf{a}_\text{TX}^\text{H}(\theta_{d,l}^u),
\end{align}
where $N_\text{cluster}$ is the number of clusters, $N_\text{path}$ is the number of paths in each cluster, $g_{d,l}^{u}=\sigma_d^u \sqrt{\kappa_{d,l}^{u}} e^{j \varphi_{d,l}^{u}}$ is the coefficient of the path $l$ in cluster $d$ in which $\sigma_d^u$ models path-loss and shadowing, $\kappa_{d,l}^{u}$ models the fraction of the signal power carried by the $l$-th path within the $d$-th cluster and $\varphi_{d,l}^{u} \sim \mathcal{U}(0,2\pi)$ models the phase of the $l$-th path within the $d$-th cluster, $\tau_{d,l}^u={\tau_0}_{d}^u+{\tau_1}_{d,l}^u$ represents the delay of the path $l$ in cluster $d$ in which ${\tau_0}_{d}^u$ is the group delay of the $d$-th cluster and ${\tau_1}_{d,l}^u$ is the relative path delay of the $l$-th path in the $d$-th cluster, $f_q$ is the center frequency of block $q$, $\mathbf{a}_\text{TX}(.)$ and $\mathbf{a}_\text{RX}(.)$ are the array response vectors at the BS and UE $u$, and finally $\theta_{d,l}^u$ and $\phi_{d,l}^u$ represent the angle of arrival (AoA) and the angle of departure (AoD) corresponding to the $l$-th path in cluster $d$ of the channel. The generation of these channel parameters follows Section IV-F in \cite{DL}. Based on $\mathbf{G}_{q,u}$, BA and CE are performed as follows. 
Note that $g_{q,n,u,m}^\text{eff}=\left({\mathbf{v}_{l_u}}\right)^H \mathbf{G}_{q,u}\left(\mathbf{w}_{h(u,m)}\right)$. Recall that $g_{q,n,u,m}^\text{eff}$ represents the effective channel gain between UE $u$ and the BS using $\mathbf{w}_{h(u,m)}$ if $n = u$ or the effective channel interference between UE $n$ and the BS due to the transmission of UE $u$ to the BS using $\mathbf{w}_{h(n,m)}$ if $n \neq u$, for $\forall m \in \{1,...,M\}$.

We consider a low-mobility mmWave scenario with pedestrian UEs. We assume the UEs are fixed and the large-scale fading $\sigma_d^u$ remains constant over $T$ time slots and that BA is performed once at the beginning of these $T$ time slots to capture the large-scale fading characteristics. Specifically, for each UE, all combinations of a BS beam and a UE beam are evaluated, and the $M$ pairs that best compensate for the large-scale fading (i.e., yield the $M$ highest effective channel gains) are selected for that UE \cite{ECSIAhmed,BAGain}. 

We assume one CE block consists of $N_F=6$ channels in frequency, corresponding to $Q = 22$ blocks in the whole bandwidth. CE is performed $Q$ times per time slot for each block. The value of \( Q \) was selected based on the channel model, to ensure that the bandwidth of each block is smaller than the coherence bandwidth. Specifically, we generated \( 5 \times 10^4 \) \emph{channel samples} by simulating 50 realizations, each with 10 users over 100 time slots, and computed the average root mean square (RMS) delay \( \bar{\tau}_\text{RMS} \) across all channel samples. \( N_F \) is set such that $ \frac{BW}{Q} \ll \frac{1}{\bar{\tau}_\text{rms}} $. For instance, we used \( \frac{BW}{Q} \approx 0.1 \frac{1}{\bar{\tau}_\text{RMS}}   \) to satisfy this condition \cite{DL}.

\subsection{Evaluation Criterion}
We define a network realization $\gamma(U)$ as the uniform placement of $U$ users in the cell where, for each UE,  the MIMO channel parameters for each $M$ pairs of beams where the UE beam is fixed for each UE and there are $M$ matched BS beams per UE are given for each $Q$ blocks for $T=100$ time slots.  For each realization $\gamma(U)$, RRM is performed for each of the $T$ time slots to obtain $GM(\gamma(U))$ (see Eq.~(\ref{GMGM}) in Section~\ref{FN}). To evaluate the performance for a given value of $U$, we generate $Z=50$ realizations and let $\overline{GM}(U) = \frac{1}{Z} \sum_{z=1}^Z GM(\gamma_z(U))$ denote the arithmetic mean of the GM of each realization. 

Finally, the moving average parameter $W$ is set to 10. The initial value of $R_u$ is set to 2 for all users. 

\subsection{Numerical Evaluations for Joint RRM Optimization}

\begin{figure}
    \centering
    \includegraphics[width=85mm]{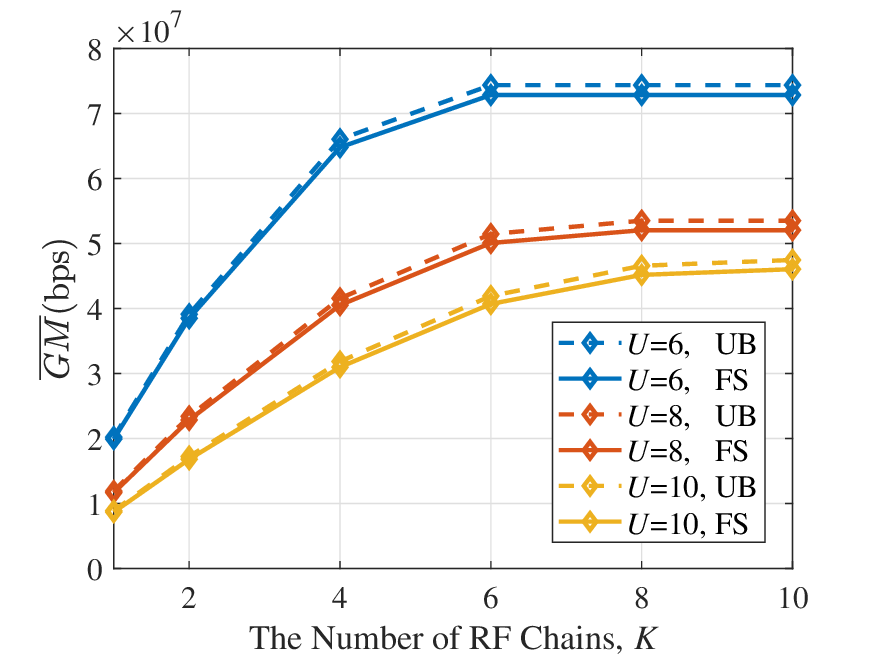}
    \caption{Quality of the feasible solution for $U=6,8,10$ and $M=1$ (UB: upper bound, FS: feasible solution).}
    \label{r1.1}
    \vspace{-5pt}
\end{figure}

We start by analyzing the quality of our feasible solution to the joint RRM optimization Problem $\Omega$. Fig.~\ref{r1.1} compares the performance of the feasible solution proposed in Section~\ref{FS} with the upper bound presented in Section~\ref{UpperBound} as a function of the number of BS RF chains $K$. The gap between the feasible and upper-bound performances is small, at most 3\% of the upper bound, confirming the quasi-optimality of the proposed feasible solution. Moreover, the performance begins to saturate way before reaching $K = U$. For instance, when $U=10$, setting $K=6$ already achieves 88\% of the performance observed at $K=U$. We will further investigate the impact of $K$ using the offline heuristic in scenarios with larger values of $U$.

\begin{figure}
    \centering
    \includegraphics[width=85mm]{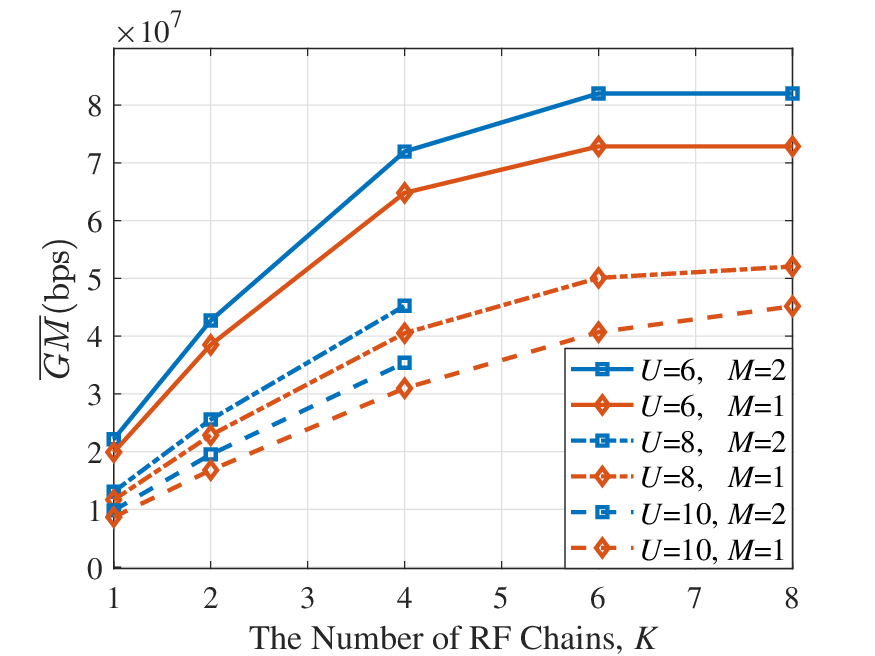}
    \caption{Impact of the number, $M$, of matched BS beams per UE on the performance of the feasible solution for $U=6,8,10$.}
    \label{r1.2}
    \vspace{-5pt}
\end{figure}

Next, we analyze the impact of the number, $M$, of matched BS beams per UE  on the performance of  the feasible solution. Fig.~\ref{r1.2} depicts the system performance as a function of $K$ for $M=1$ and $M=2$. We observe a maximum gain of 17\% at $K=2$ and $U=10$ by using $M=2$ instead of $M=1$. With larger $K$ (e.g., $K=6$ and $U=6$), doubling $M$ can increase the performance by 12\%. Recall that this performance improvement is at the cost of increased channel estimation overhead. Since increasing $U$, $K$ and $M$  substantially increases the number of variables in the joint optimization, solving the problem becomes intractable for $U>10$, $K>4$ and $M>2$. Therefore, we will explore the impact of $M$ with larger $U$ and $K$ using the offline heuristic in the next subsection.

\subsection{Numerical Evaluation of the Offline Heuristic}
\begin{figure}
    \centering
    \includegraphics[width=85mm]{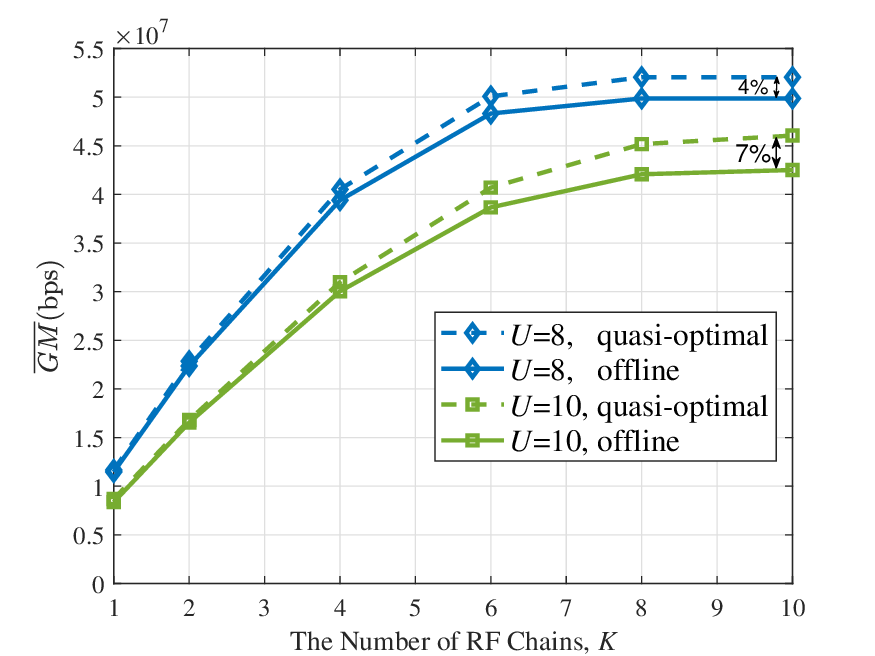}
    \caption{Performance of the offline heuristic as a function of $K$ for $U=8,10$ and $M=1$.}
    \label{r2.1}
    \vspace{-5pt}
\end{figure}

We first validate our offline heuristic by comparing it to the quasi-optimal solution. Fig.~\ref{r2.1} presents the performance of our offline heuristic as a function of $K$, alongside the target performance obtained from the quasi-optimal solution. Our offline heuristic achieves at least 96\% of the target performance for $U=8$ and 93\% for $U=10$. Notably, the runtime for obtaining the heuristic curve with $U=10$ is one hour and 46 minutes, compared to the 38 days required for computing the corresponding target performance via joint RRM optimization. The performance and computational efficiency of the offline heuristic allow us to obtain engineering insights for larger systems. 
This makes the proposed offline heuristic well-suited for system planning in large-scale deployments.

\begin{figure}
    \centering
    \includegraphics[width=85mm]{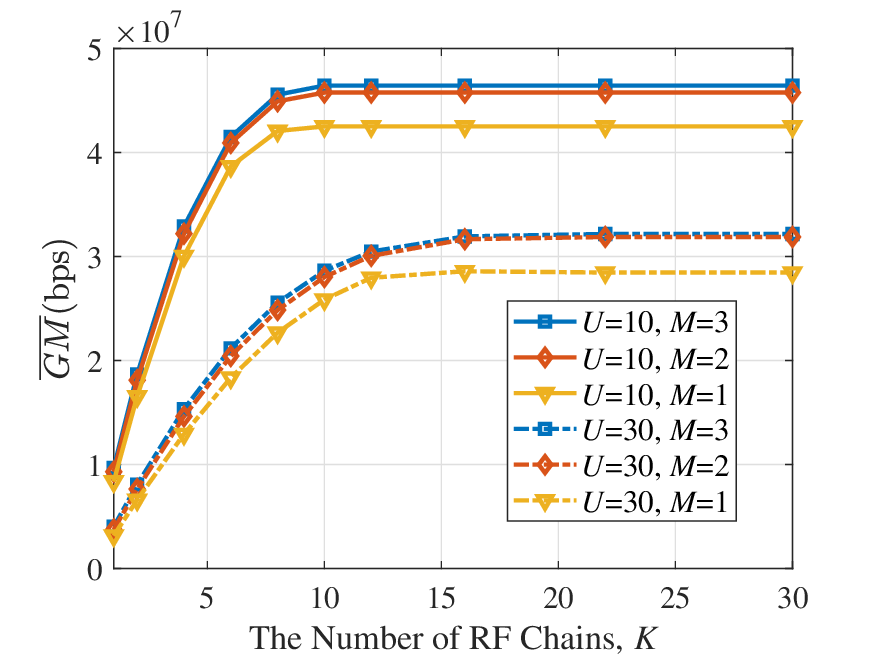}
    \caption{Impact of the number of matched BS beams per UE on offline heuristic performance for $U=10$  and $30$.}
    \label{r2.2}
    \vspace{-5pt}
\end{figure}

With our offline heuristic, we can investigate the impact of $M$ under large-scale scenarios with large $U$ and $K$. Fig.~\ref{r2.2} presents the system performance as a function of $K$ for configurations with up to $U=30$, $K=30$, and $M=3$. The maximum improvement obtained by increasing $M=1$ to $M=2$ is 19\% at $U=30$ and $K=1$. The performance gain diminishes to 12\% as $K$ increases to 30 for $U=30$ and 8\% for $U=10$ and $K=10$. When $K$ is small, only a limited number of UEs can be scheduled within a time slot. Increasing $M$ allows more UEs to be selected, thereby enabling more effective utilization of uplink power. As a result, the performance gain from increasing $M$ is more significant when $K$ is small. However, when comparing $M=2$ with $M=3$, the maximum observed gain is 7\% at $U=30$ and $K=1$, while the improvement becomes negligible as $K$ increases for both $U=10$ and $U=30$. This is because the third (and subsequent) best-matched BS beams typically offer much weaker channel gains for a UE than the first two and thus contribute minimally to performance, even with additional flexibility in user selection. 

Another observation from Fig.~\ref{r2.2} is that for $U=30$, setting $K=10$ achieves 90\% of the performance at $K=U$ when $M=1$ and 88\% when $M=2$ or $M=3$. It further validates that good performance can be achieved even when $K$ is significantly smaller than $U$. This is because, as $K$ increases, allowing more UEs within the same PRB, the channel gains after ZF-DBF tend to decrease due to reduced spatial separation. This degradation leads to lower per-UE rates, thereby diminishing the marginal performance gains from further increasing $K$.

\subsection{Numerical Evaluations of the Online Heuristic}
\begin{figure}
    \centering
    \includegraphics[width=85mm]{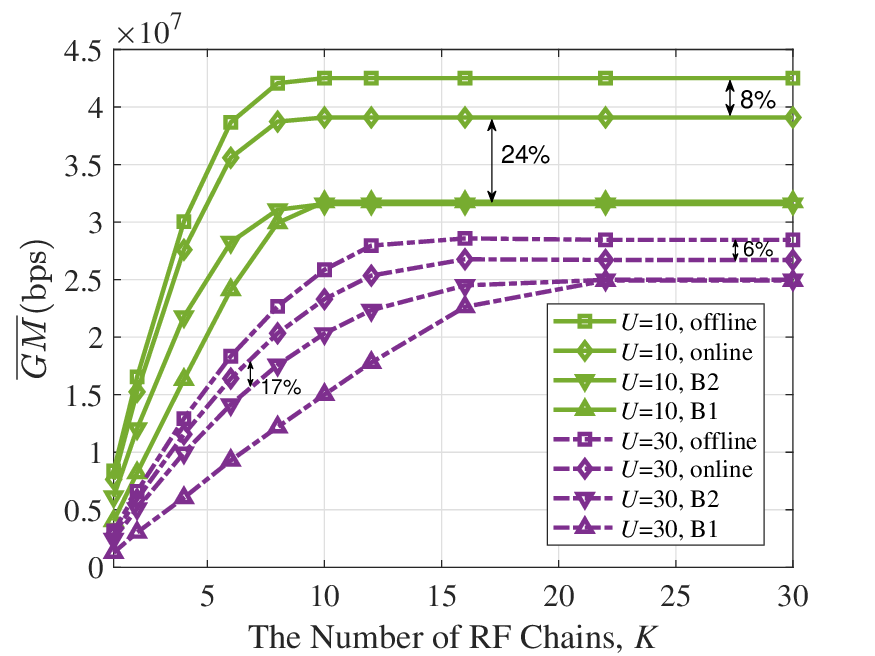}
    \caption{Performance of the online heuristic as a function of $K$ for $U=10$  and $30$, and $M=1$.}
    \label{r3.1}
    \vspace{-5pt}
\end{figure}

We begin by comparing the performance of the online heuristic against that of the offline heuristic. We set the parameter $X$ (number of extra steps in the search) of PBS to  $X=N_F+1=7$. Fig.~\ref{r3.1} shows the performance of the online heuristic as a function of $K$ for $U=10$ and $U=30$, along with the offline heuristic as the target. The performance of our online heuristic is at least 92\% of the target performance for $U=10$ and at least 94\% for $U=30$. The performance gap primarily arises from the use of WFPA in the online heuristic, as opposed to the optimization employed in the offline heuristic. Although WF does not achieve a similar performance as PA optimization, its millisecond-level runtime makes it suitable for operation. 

Fig.~\ref{r3.1} also compares the proposed online heuristic with two benchmarks. B1 uses round-robin beam selection, the scheduler from \cite{PerBeam} for per-beam user selection,  ZF-DBF, and WFPA. B2 uses the load-aware beam selection proposed in our heuristic, the scheduler from \cite{PerBeam} for per-beam user selection,  ZF-DBF, and WFPA. Recall that our online heuristic employs load-aware beam selection, our proposed PBS,  ZF-DBF, and WFPA. Specifically, B1 is used as the baseline to demonstrate the effectiveness of our online heuristic. B2 is intended to show the performance improvement brought by PBS compared with the scheduler from \cite{PerBeam}. Comparing our online heuristic with B2 in Fig.~\ref{r3.1}, we see that employing PBS instead of the scheduler in \cite{PerBeam} improves performance by up to 24\% for $U=10$ and up to 17\% for $U=30$. Comparing B2 with B1, we see that load-aware beam selection brings a significant performance boost (up to 52\% for $U=10$ and up to 93\% for $U=30$), especially when $K$ is small.

\begin{figure}
    \centering
    \includegraphics[width=85mm]{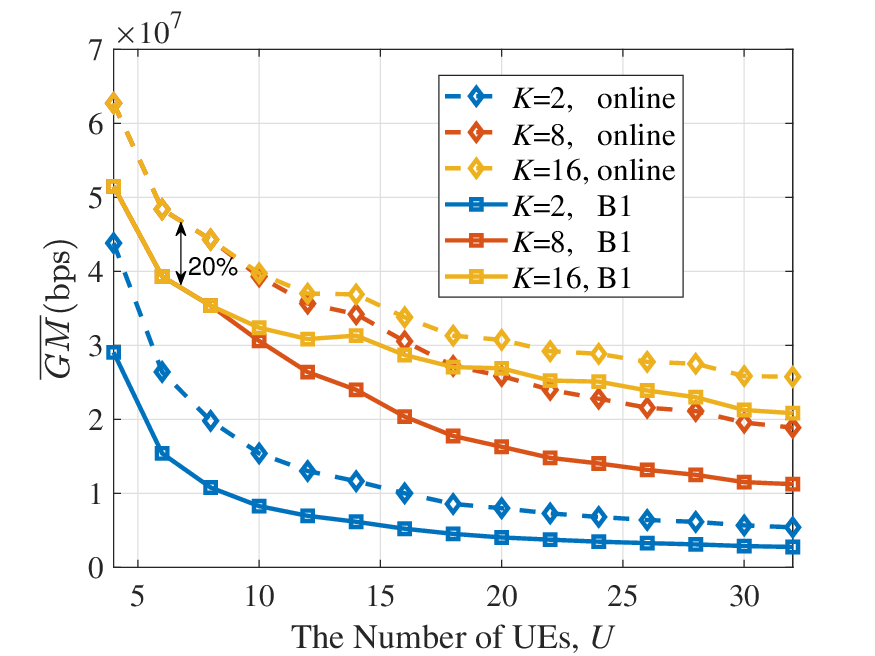}
    \caption{Performance comparison between the online heuristic and a baseline benchmark as a function of $U$ for $K=2,8,16$ and $M=1$.}
    \label{r3.2a}
    \vspace{-5pt}
\end{figure}

To provide a more comprehensive comparison, we further evaluate our proposed online heuristic against B1 in Fig.~\ref{r3.2a} with fixed values of $K$. Note that B2 is not included in this figure because it incorporates our proposed beam selection method and therefore is not considered a baseline. Comparing against B1 offers a clearer illustration of the overall advantages of our proposed online heuristic. As shown in Fig.~\ref{r3.2a}, our online heuristic outperforms the benchmark by 23-24\% for $K=16$, 23-53\% for $K=8$ and 51-97\% for $K=2$, which demonstrate the effectiveness of our online heuristic.

\begin{figure}
    \centering
    \includegraphics[width=85mm]{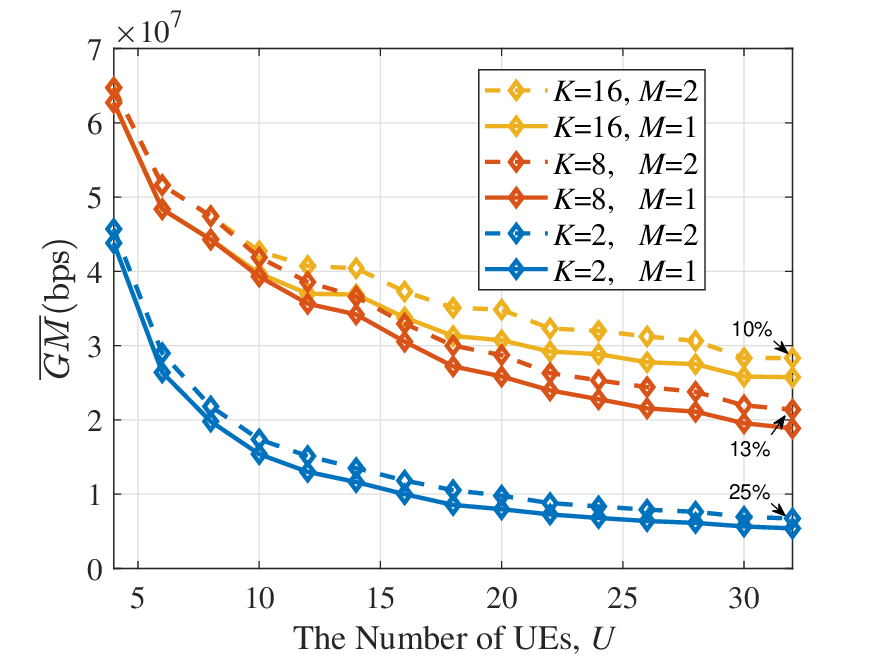}
    \caption{Performance of the online heuristic as a function of $U$ for $K=2,8,16$ and $M=1,2$.}
    \label{r3.2}
    \vspace{-5pt}
\end{figure}

We further evaluate our online heuristic with fixed values of $K$ and $M$, examining its performance as a function of the number of users, $U$. Based on insights from the offline heuristic results, we consider $K = 2, 8, 16$ and $M = 1, 2$. The number of realizations is increased to $Z = 200$ since the online heuristic is much faster than the offline solution. As illustrated by Fig.~\ref{r3.2}, $K$ continues to play a critical role in determining system performance when $U$ is large. For instance, at $U=32$ and $M=1$, increasing $K$ from 2 to 8 nearly triples the system performance. While the performance gap between $K=8$ and $K=16$ remains under 10\% for $U<16$, it grows significantly with higher $U$, reaching 33\% for $M=1$ and 37\% for $M=2$ when $U=32$. Fig.~\ref{r3.2} also shows that using $M>1$ is more beneficial when $U$ is large and $K$ is small. Specifically, at $U=32$, increasing $M$ from 1 to 2 yields gains of 25\%, 13\%, and 10\% for $K=2$, $K=8$, and $K=16$, respectively.

Finally, we present the time complexity analysis of each step of the proposed online heuristic in Table~\ref{CPLEX}. The analysis assumes parallel execution across beams for PBS, across blocks for Step~2, across PRBs within a time slot for ZF-DBF, and across UEs for WF. For each beam, PBS searches up to $Q$ blocks and at most another $N_F$ PRBs if the final block is not selected. In each search step, PBS will find the best UE among at most $U$ UEs (if all UEs are matched with this beam). Consequently, the worst-case complexity of PBS per beam is $\mathcal{O} \big((Q+N_F) U \big)$. Step~2 is performed on a per-block basis. Within each block, it checks up to $N_F$ PRBs sequentially to remove or switch repeated UEs. This results in a worst-case complexity of $\mathcal{O}(N_F)$ per block. Load-aware beam selection has a complexity of $\mathcal{O} (|\mathcal{I}_s|) $ because it finds the $L$ beams with the highest weighted sum rates from $|\mathcal{I}_s|$ beams. Recall that $L$ is the minimum between the number, $K$, of RF chains at the BS and the number, $|\mathcal{I}_s|$, of matched BS beams (e.g., $L \leq 16$ for Fig.~\ref{r3.2}). Per-PRB ZF-DBF is performed on at most $L$ UEs, hence its complexity is $\mathcal{O} (L^3) $ \cite{MCS}.  Finally, the complexity of WF is linear with respect to the number of channel coefficients \cite{WF} in PA, which, in our system, results in a worst-case complexity of $\mathcal{O}(Q)$ per UE. Among these steps, PBS and ZF-DBF dominate the overall complexity. Step~2, load-aware beam selection and WF have negligible complexity because $N_F \ll (Q+N_F) U$, $|\mathcal{I}_s| \propto U \ll (Q+N_F) U$ and $Q \ll (Q+N_F) U$. The overall complexity is $\mathcal{O} \big((Q+N_F) U + L^3 \big)$. For a given system where $K, Q$, and $N_F$ are fixed, the complexity scales linearly in $U$ since for $U$ large, $L=K$. Combined with its near-offline performance (92\% at least), the low complexity of our online heuristic shows its suitability for system operation.

\begin{table}
\centering
\setlength{\tabcolsep}{1.5mm}{
\caption{Complexity of the online heuristic solution (LA BSel: load-aware beam selection)}
\label{CPLEX}
\footnotesize
\begin{tabular}{|c|c|c|c|c|c|} 

\hline

\rowcolor[HTML]{D3D3D3} Step~1: PBS & Step~2 ($M>1$) & LA BSel & ZF-DBF & WF \\
\hline

$\mathcal{O} \big((Q+N_F) U \big)$ & $\mathcal{O} (N_F)$ & $\mathcal{O} (|\mathcal{I}_s|) $ & $\mathcal{O} (L^3) $ & $\mathcal{O} (Q)$ \\
\hline

\end{tabular}}

\vspace{-5pt}

\end{table}



\section{Conclusion} \label{sec:conclusion}
In this paper, we have investigated RRM for the uplink of codebook-based HBF systems with multiple channels and no assumptions on the number of RF chains at the BS. Existing per-PRB solutions proposed in the uplink literature are not applicable in our scenario, as both beam selection and power allocation (PA) must be performed at the time-slot level. Per-time-slot solutions in DL are also unsuitable because the per-UE PA in UL is inherently different from that in DL where the BS distributes power to different UEs. To address this gap, we first formulate a per-time-slot joint RRM optimization problem including beam selection, user selection and PA. We can solve this problem to quasi-optimality for a small number of UEs ($U \leq 10$). To support large system planning (e.g., $30$ UEs, $30$ RF chains at the BS and $3$ matched BS beams per UE), we propose an offline heuristic using per-beam user selection and PA optimization with a load-aware beam selection. This offline heuristic can achieve at least 93\% of the quasi-optimal performance from joint optimization while reducing the runtime from months to hours. For online deployment, we  propose an online heuristic by replacing the per-beam optimization with a proposed per-beam scheduler, which outperforms the state-of-art solution by 22\% at least, and by substituting power optimization with water filling in the last step. Our online heuristic maintains at least 92\% of the offline heuristic with very low complexity. 

We have also provided engineering insights into the impact of key system parameters, i.e.,  the number, $K$, of RF chains at the BS and the number, $M$, of matched BS beams per UE. Our results indicate that setting $K = 0.6 U$ (resp. $K=\frac{1}{3}U$) is sufficient to achieve strong performance, at least 88\% of that at $K=U$ for $U=10$ (resp. $U=30$). Furthermore, increasing $M$ from 1 to 2 can yield a substantial performance gain, up to 25\%,  when $U$ is large and $K$ is small (e.g., $U=32$ and $K=2$). However, increasing $M$ from 2 to 3 results in negligible additional improvement. As part of future work, we aim to extend our RRM solutions to multi-cell scenarios, where inter-cell interference plays an important role. A practical approach that does not require cooperation between cells is to impose transmit-power constraints on UEs, model inter-cell interference as statistical noise and use the  heuristics developed in this paper. Another important direction of future work is to develop a full-CSI-based upper bound to explore the full potential of the system.

\ifCLASSOPTIONcaptionsoff
  \newpage
\fi



%
\bibliographystyle{IEEEtran}
\bibliography{References.bib}

\begin{IEEEbiography}[{\includegraphics[width=1in,height=1.25in,clip,keepaspectratio]{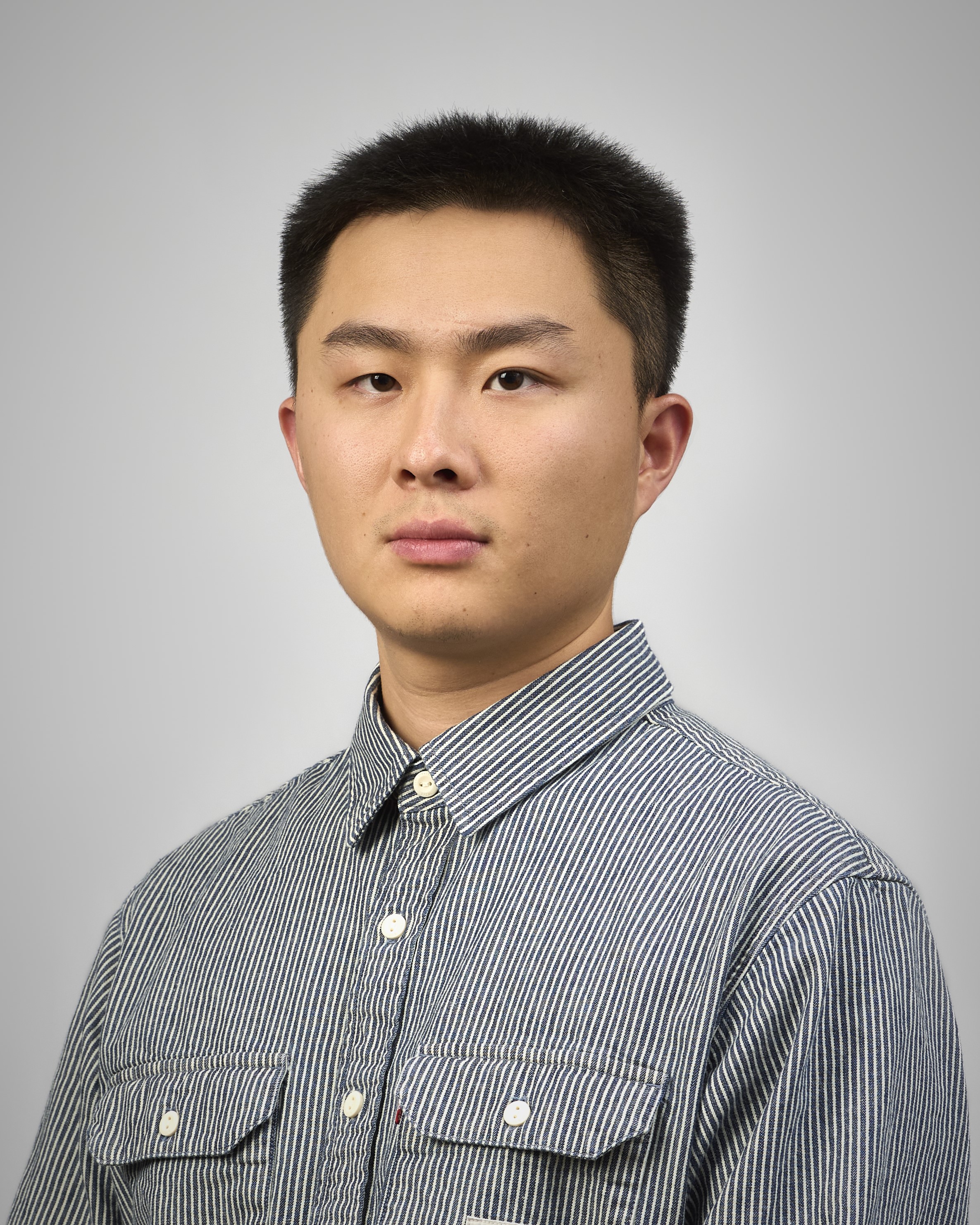}}]{Yuan Quan} received the B.Sc. degree in information and communication engineering from the Beijing Institute of Technology, Beijing, China, in 2018, the M.Sc. degree in information and communication engineering from the National University of Defense Technology, Changsha, Hunan, China, in 2020, and the Ph.D. degree in electrical and computer engineering from the University of Waterloo, Waterloo, ON, Canada, in 2025. He is currently working as a system engineer at BLiNQ Networks in Markham, ON, Canada. His research interests include wireless networks, optimization, resource management, and signal processing.
\end{IEEEbiography}

\begin{IEEEbiography}[{\includegraphics[width=1in,height=1.25in,clip,keepaspectratio]{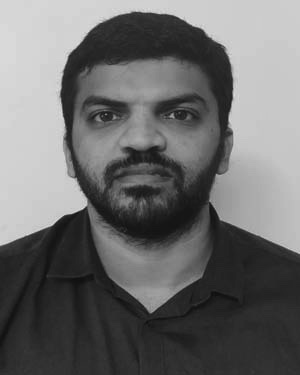}}]{Haseen Rahman} received the B.Tech degree from the Government College of Engineering Kannur, India, master’s degree in communication engineering from IIT Madras in 2012 and Ph.D degree with the Department of Electrical Engineering, IIT Bombay,Mumbai, India in 2021. He was a post doctoral fellow at University of Waterloo, ON, Canada and is currently working as wireless researcher at Skycope Technologies in BC, Canada. His research interests include resource allocation in wireless communication and network information theory.
\end{IEEEbiography}

\begin{IEEEbiography}[{\includegraphics[width=1in,height=1.25in,clip,keepaspectratio]{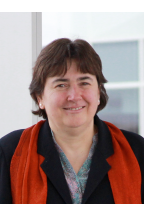}}]
{Catherine Rosenberg} (M'89-SM'96-F'11) is an Emeritus Professor with the Department of Electrical and Computer Engineering, University of Waterloo, Waterloo, ON, Canada. Her research interests are in networking, wireless and energy systems. More information is available at https://catherinerosenberg.com/
\end{IEEEbiography}

\end{document}